\documentclass[twocolumn,traditabstract,longauth]{aa} 

\usepackage[hyperindex,breaklinks=true]{hyperref}
\hypersetup{
  colorlinks   = true, 
  urlcolor     = blue,
  linkcolor    = blue, 
  citecolor    = blue  
}   
\usepackage{natbib,twoopt}                  
\bibpunct{(}{)}{;}{a}{}{,}                 

\usepackage{graphicx}
\usepackage{tablefootnote}
\usepackage[binary-units]{siunitx}
\usepackage{txfonts}
\usepackage{xcolor}
\usepackage{multirow}
\usepackage{rotating}
\usepackage{orcidlink}
\usepackage{lscape}
\usepackage{subcaption}

\usepackage{amsmath}
\usepackage{amssymb}
\usepackage[T1]{fontenc}
\usepackage[utf8]{inputenc}
\usepackage{bm}
\usepackage{ragged2e}

\begin{document} 

   \title{Planet formation in chemically diverse and evolving discs}

   \subtitle{I. Composition of planetary building blocks}

   \author{E. Pacetti\inst{1}\fnmsep\thanks{Corresponding author}\and
          E. Schisano\inst{1}\and 
          D. Turrini\inst{2}\and 
          C. P. Dullemond\inst{3}\and 
          S. Molinari\inst{1}\and
          C. Walsh\inst{4}\and
          S. Fonte\inst{1}\and 
          U. Lebreuilly\inst{5}\and
          R. S. Klessen\inst{3,6,7,8}\and
          P. Hennebelle\inst{5}\and
          S. L. Ivanovski\inst{9}\and
          R. Politi\inst{1}\and
          D. Polychroni\inst{2,9,10}\and
          P. Simonetti\inst{9,10}\and
          L. Testi\inst{11,12}}

   \institute{INAF Istituto di Astrofisica e Planetologia Spaziali, Via Fosso del Cavaliere 100, I-00133, Roma, Italy\\
              \email{elenia.pacetti@inaf.it}\and
             INAF Osservatorio Astrofisico di Torino, Via Osservatorio 20, I-10025, Pino Torinese, Italy\and
            Institut für Theortische Astrophysik, Universität Heidelberg, Zentrum für Astronomie, Albert-Ueberle-Str 2, D-69120 Heidelberg, Germany\and
            School of Physics and Astronomy, University of Leeds, Woodhouse Lane, Leeds, LS2 9JT, UK\and
            Université Paris-Saclay, Université Paris Cité, CEA, CNRS, AIM, 91191, Gif-sur-Yvette, France\and
            Interdisziplinäres Zentrum für Wissenschaftliches Rechnen, Universität Heidelberg, Im Neuenheimer Feld 225, D-69120 Heidelberg, Germany\and
            Harvard-Smithsonian Center for Astrophysics, 60 Garden Street, Cambridge, MA 02138, USA\and
            Elizabeth S. and Richard M. Cashin Fellow at the Radcliffe Institute for Advanced Studies at Harvard University, 10 Garden Street, Cambridge, MA 02138, USA\and
            INAF – Osservatorio Astronomico di Trieste, via G. B. Tiepolo 11, I-34143, Trieste, Italy\and
            ICSC – National Research Centre for High Performance Computing, Big Data and Quantum Computing, Via Magnanelli 2, I-40033, Casalecchio di Reno, Italy\and
            Dipartimento di Fisica e Astronomia "Augusto Righi", Viale Berti Pichat 6/2, Bologna, Italy\and
            INAF – Osservatorio Astrofisico di Arcetri, Largo E. Fermi 5, 50125, Firenze, Italy
             }

   \date{Received February 03, 2025; accepted June 20, 2025}

    \abstract{Protoplanetary discs are dynamic environments where the interplay between chemical processes and mass transport shapes the composition of gas and dust available for planet formation. We investigate the combined effects of volatile chemistry - including both gas-phase and surface reactions - viscous gas evolution, and radial dust drift on the composition of planetary building blocks. We explore scenarios of chemical inheritance and reset under varying ionisation conditions and for various dust grain sizes in the sub-mm regime. We simulate disc evolution using a semi-analytical 1D model that integrates chemical kinetics with gas and dust transport, accounting for viscous heating, turbulent mixing, and refractory organic carbon erosion. We find that mass transport plays a role in the chemical evolution of even sub-\si{\micro\metre} grains, especially in discs that have experienced strong heating or are exposed to relatively high levels of ionising radiation. The radial drift of relatively small ($\sim100$~\si{\micro\metre}) icy grains can yield significant volatile enrichment in the gas phase within the snowlines, increasing the abundances of species like H$_2$O, CO$_2$, and NH$_3$ by up to an order of magnitude. Early planetesimal formation can lead to volatile depletion in the inner disc on timescales shorter than 0.5~Myr, while the erosion of refractory organic carbon can lead to markedly superstellar gas-phase C/O and C/N ratios. Notably, none of the analysed scenarios reproduce the classical monotonic radial trend of the gas-phase C/O ratio predicted by early models. Our results also show that a pairwise comparison of elemental ratios, in the context of the host star’s composition, is key to isolating signatures of different scenarios in specific regions of the disc. We conclude that accurate models of planet formation must concurrently account for the chemical and dynamical evolution of discs, as well as the possible diversity of their initial chemical and physical conditions.}

   \keywords{Protoplanetary discs -- Astrochemistry -- Planetary systems -- Planets and satellites: formation, composition}

   \maketitle
%

\section{Introduction}\label{sec:intro}

The remarkable advances in observational astronomy over the past decade have profoundly transformed our understanding of the conditions and processes driving planet formation in protoplanetary discs. The morphology and composition of discs have been characterised down to scales of less than 10~au, revealing a high occurrence of substructures and rich gas- and ice-phase chemistry \citep[e.g.][]{Andrews2018, Oberg2021b, McGuire2022, Kamp2023, McClure2023, Sturm2023, Oberg2023, Manara2023}. Observations also include discs hosting confirmed protoplanets in the final stages of their formation \citep[e.g.][]{Perotti2023}. Simultaneously, detailed spectroscopic characterisations of planetary atmospheres across a large sample of exoplanets have revealed the presence of both volatile and refractory species \citep[e.g.][]{Giacobbe2021, Changeat2022, Guilluy2022, Carleo2022, Tsai2023, Edwards2023}.

Access to the chemical composition of both protoplanetary discs and planetary atmospheres allows direct comparison of their elemental abundance ratios, which has long been recognised as a key diagnostic for planet formation processes \citep[see][for a review]{Madhusudhan2019}. Since the seminal work by \citet{Oberg2011}, the carbon-to-oxygen (C/O) ratio has received particular attention. In discs, thermal freeze-out drives phase transitions at the snowlines of the major volatile carriers of oxygen and carbon (H$_2$O, CO$_2$, and CO), resulting in radial variations in the C/O ratio of both gas and solids. These variations have motivated the development of a diagnostic framework to link the C/O ratios in planetary atmospheres to the formation locations of planets within the disc \citep[e.g.][]{Madhusudhan2014, Madhusudhan2017, Mordasini2016}. However, recent models have demonstrated that early implementations of this approach, while conceptually insightful, tend to oversimplify the complex physical, dynamical, and chemical evolution of protoplanetary discs suggested by observations \citep[e.g.][]{Bergin2016, Krijt2020, Bosman2021, Bergner2024}.

Beyond temperature, other key parameters influencing disc chemistry include density, ionisation rate, and chemical initial conditions \citep[e.g.][]{Henning2013}. Variations in these factors can drive the conversion of abundant species into molecules with different volatilities, altering the elemental ratios of gas and solids within the disc. A central unresolved question is whether discs inherit the molecular composition of their parent molecular cloud or undergo a chemical reset during collapse and disc formation \citep[e.g.][]{Oberg2021}. Each scenario leads to significantly different chemical outcomes, with direct effects on both disc chemistry \citep{Eistrup2016, Eistrup2018} and the composition of planetary atmospheres \citep{Pacetti2022}.

Volatiles, however, account for only part of the picture. A growing body of evidence from the interstellar medium (ISM), comets, meteorites, circumstellar discs, and polluted white dwarfs points to a more complex partitioning of elements, involving significant contributions from refractory and semi-refractory phases \citep[e.g.][]{Bardyn2017, Altwegg2020}. A long-standing issue in Solar System formation is the observed deficit of solid carbon, particularly in the Earth and the inner Solar System \citep{Lee2010, Li2021}. This “missing carbon” problem can be explained by recognising that up to half of the cosmic carbon budget is sequestered in semi-refractory organic materials \citep{Gail2017, vantHoff2020}. These compounds, embedded in icy grains, can be transported inwards and thermally processed in the inner disc, where they enrich the gas phase upon sublimation. Cometary C/Si ratios comparable to those of interstellar dust \citep{Savage1996, Rubin2019} suggest that these organics are inherited from earlier evolutionary stages. In contrast, meteorite classes (carbonaceous, ordinary, and enstatite chondrites) exhibit a systematic decline in carbon content with decreasing formation distance from the Sun \citep{Allegre2001, Bergin2015}, consistent with the progressive erosion of refractory organics in the inner disc. Similar trends are observed in exoplanetary systems: polluted white dwarfs display strongly subsolar C/Fe and C/Si ratios, indicative of the accretion of carbon-poor, asteroidal material \citep{Jura2006, Jura2012, Gansicke2012, Farihi2016}. Together, these findings suggest that the redistribution and erosion of refractory carbon is a common outcome of disc evolution and planet formation \citep{Nazari2023}.

These considerations are particularly relevant given that exoplanet formation is often modelled assuming solar-like native environments. However, homogeneous spectroscopic surveys of stars in the solar neighbourhood reveal significant compositional diversity, with trends in stellar metallicity and elemental ratios \citep[e.g.][]{Magrini2022, Biazzo2022, daSilva2024, Filomeno2024}. Consequently, the Sun does not necessarily provide a representative chemical template for all planetary systems \citep{Teske2024}, emphasising the importance of interpreting disc and exoplanet compositions in the context of their specific stellar environment \citep{Turrini2021, Pacetti2022}.

Finally, disc dynamics plays a critical role in shaping the spatial and temporal distribution of chemical species. Protoplanetary discs serve as mass transport channels, with gas accreting onto the central star and dust evolving under the influence of gas drag, radial drift, and turbulent mixing \citep{Lynden-Bell1974, Shakura1973, Weidenschilling1977a, Youdin2007, Birnstiel2010}. These processes shift the location of snowlines and influence volatile delivery to the inner disc \citep[e.g.][]{Dodson2009, Piso2015}. In particular, radial drift of icy pebbles has been extensively studied for millimetre- to centimetre-sized pebbles and has been shown to enhance gas-phase volatile abundances by releasing ices near snowlines \citep[e.g.][]{Cuzzi2004, Oberg2016, Booth2017, Booth2019, Schneider2021a, Schneider2021b}. Observational evidence for such enhancements has emerged in multiple discs, and has been attributed to pebble drift \citep[e.g.][]{Zhang2019, Banzatti2020, Banzatti2023, Perotti2023, Mah2023, Schwarz2024}, the erosion of refractory organic carbon \citep[e.g.][]{Tabone2023, Kanwar2024}, or trapping mechanisms beyond snowlines \citep[e.g.][]{vanderMarel2021}.

Disc evolution can reshape the composition of planet-forming material on timescales of a few million years, comparable to those of planet formation \citep[e.g.][]{Eistrup2018, Booth2019}. Models incorporating more realistic treatments of disc chemistry have shown that the planetary C/O ratio cannot always be unambiguously linked to a specific formation scenario \citep[e.g.][]{Cridland2019a, Turrini2021, Pacetti2022}. This degeneracy can be partially resolved by adopting a multi-element approach that includes elements with a range of volatilities, which provide complementary constraints on disc chemistry and planet formation pathways \citep[e.g.][]{Oberg2019, Cridland2017b, Cridland2020, Schneider2021b, Turrini2021, Pacetti2022, Chachan2023, Crossfield2023, Danti2023}. For example, \citet{Turrini2021} investigated fingerprints of planet formation in the atmospheric abundances of C, O, N, and S in giant planets. Their study assumed a stationary disc structure but incorporated a detailed compositional model with volatiles inherited from the molecular cloud and refractory materials calibrated using data from the Solar System and extrasolar systems. In \citet{Pacetti2022}, we extended this framework by exploring the effects of different initial chemical conditions in the disc, including scenarios of chemical reset and exposure to ionising radiation.

This paper is the first in a two-part series that builds on the work of \citet{Turrini2021} and \citet{Pacetti2022} to investigate the effects of disc evolution on the composition of planet-forming material and the atmospheres of giant planets. In this study, we focus on the disc itself, exploring how different initial chemical conditions and dust grain sizes influence the abundances of volatile molecules and the C/O, C/N, and N/O ratios. Compared to previous models of coupled chemical and dynamical evolution \citep[e.g.][]{Eistrup2018, Cridland2017b, Booth2019, Soto2022}, our work introduces two key extensions: a more realistic treatment of the initial volatile composition and its evolution - including the erosion of semi-refractory organic carbon - and the inclusion of planetesimal formation as a sink for volatiles. We specifically examine submillimetre-sized grains, considering a range of grain sizes smaller than those typically assumed for pebbles in discs.

We introduce our time-dependent disc model and its numerical implementation in Sect.~\ref{sec:methods}. We then present our results for a stationary disc, where we consider chemical processes but exclude mass transport (Sect.~\ref{sec:res_static_model}). Next, we examine the effects of different grain sizes on the radial distribution of gas and dust in the midplane (Sect.~\ref{sec:res_transport}) before exploring the chemical outcomes of our full model under various gas-dust coupling regimes and chemical scenarios (Sect.~\ref{sec:res_volatiles}). Finally, we analyse how disc evolution shapes the elemental ratios of carbon, oxygen, and nitrogen (Sects.~\ref{sec:res_ratios} and \ref{sec:res_vermetti}).


\section{Methods}\label{sec:methods}

\begin{figure}
\centering
\includegraphics[width = 0.7\columnwidth]{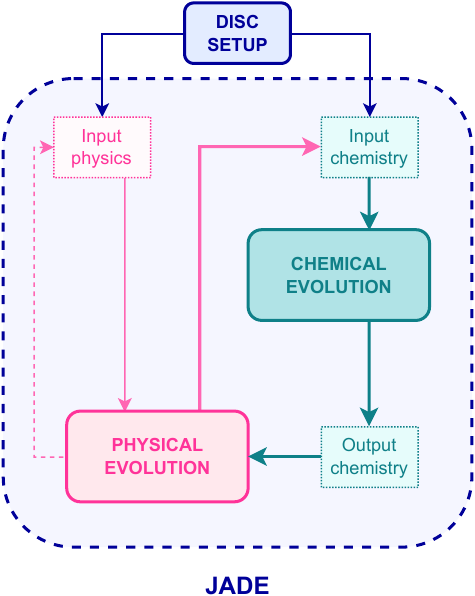}
\caption{Flowchart showing the initialisation of {\sc JADE} (\emph{Disc Setup}) and the operations performed in a single iteration of the code, following an operator-splitting scheme.} 
\label{fig:JADE}
\end{figure}

We model the planet formation environment by focusing on the midplane ($z=0$) of a Class II protoplanetary disc consisting of gas, dust, and planetesimals. The disc evolves due to the combined effects of chemical interactions and mass transport. The physical (Sect.~\ref{sec:phys_model}) and chemical (Sect.~\ref{sec:chem_model}) models of the disc are numerically integrated into our new code {\sc JADE} (Joint Astrochemistry and Disc Evolution). The code uses an operator-splitting scheme to evolve the disc, solving chemistry and physics sequentially at each time step (Fig.~\ref{fig:JADE}).

We employ {\sc JADE} to investigate how the composition of the planet-forming material changes when disc evolution is considered. We explore four chemical scenarios (two sets of initial abundances and two ionisation levels) and three populations of dust grains, resulting in twelve simulations that run for 3~Myr, with a fixed time step of $dt=10^3$~yr for both the physical and chemical modules. We note that while some chemical reactions in the disc midplane may occur on shorter timescales \citep[e.g.][]{Semenov2011}, numerical experiments with smaller time steps down to 10 yr in the scenario with the most efficient dust transport (100~\si{\micro\metre}-sized grains) showed that differences in the gas-phase abundances of key volatile carriers remain below $10\%$. This confirms that the adopted time step of $10^3$~yr does not significantly affect the global abundance trends nor the conclusions of this work.

\subsection{Disc physical model}\label{sec:phys_model}

We consider an axisymmetric, geometrically thin ($z\ll~r$), and non-self-gravitating disc accreting onto a star with solar mass, radius, and luminosity. The model accounts for viscous gas spreading as well as the effects of gas drag, radial drift, and turbulent mixing on the dust's dynamical evolution. The midplane temperature is set by both viscous heating and stellar irradiation. Further details are provided in the following subsections.  

The initial disc structure (Sect.~\ref{sec:disc_struct}) and its evolution (Sect.~\ref{sec:disc_evol}) are modelled using the {\sc DISKLAB} package\footnote{Access to this private package is available upon request to C.P. Dullemond: dullemond@uni-heidelberg.de}. We set up the disc on a 1D radial grid consisting of 600 logarithmically spaced points from $10^{-2}$~au up to $10^6$~au to avoid numerical instabilities, with 300 grid points sampling the first 100~au.

\subsubsection{Initial disc structure}\label{sec:disc_struct}

We initialise the gas surface density profile using the analytical solution for a viscous disc by \citet{Lynden-Bell1974}:
\begin{equation}
    \Sigma_{\rm g}(r)=\Sigma_{\rm c}\left(\dfrac{r}{R_{\rm c}}\right)^{-\gamma}\exp\left[-\left(\dfrac{r}{R_{\rm c}}\right)^{\left( 2-\gamma \right)}\right], 
    \label{eq:sigma_gas_paperII}
\end{equation}
where we set $R_{\rm c}=165$~au and $\gamma=0.8$ based on the observed gas surface density profile of the disc surrounding HD 163296, one of the best-characterised protoplanetary discs to date \citep{Isella2016}. The normalisation constant, $\Sigma_{\rm c}=3.3835$~\si{\gram\per\square\centi\meter}, is chosen to yield a total disc mass of 0.054~$M_\odot$, consistent with the mass of a Minimum Mass Solar Nebula (MMSN, \citealp{Hayashi1981}) with a similar radial extent \citep{Turrini2021, Pacetti2022}. 

$R_{\rm c}$ values larger than $\sim100-150$~au align with expectations for viscously evolving discs \citep[e.g.][]{Rosotti2019} and are supported by ALMA observations. Specifically, millimetre continuum data suggest dust disc radii between 10 and 500~au in a sample of approximately 200 Class II discs \citep[][and references therein]{Andrews2020}, while CO emission lines reveal gas extending up to 500~au in 22 Lupus discs \citep{Ansdell2018}. 

We assume that the disc is non-self-gravitating and has a vertically isothermal structure in hydrostatic equilibrium, following the descriptions by \citet{Nakamoto1994}, \citet{Hueso2005}, and \citet{Birnstiel2010}. Under these assumptions, the speed of sound, $c_s$, does not vary with height above the midplane (see also Sec. \ref{sec:Tmid}) and the gas surface density is related to the midplane number density of gas particles, $n_{\rm mid}$, by:
\begin{equation}
    n_{\rm mid}(r) = \dfrac{\Sigma_{\rm g}(r)}{\sqrt{2\pi}\,H_{\rm p}(r,T)\,\mu_{\rm g}\,m_{\rm p}},
    \label{eq:sigma_conversion}
\end{equation}
where $H_{\rm p}$ is the pressure scale height of the disc, defined as $H_p = c_s/\Omega_{\rm k}$, with $\Omega_{\rm k} = \sqrt{GM_\star/r^3}$, where $G$ is the gravitational constant and $M_\star$ is the mass of the central star. The quantity $\mu_{\rm g}$ is the mean molecular weight of the gas in units of the proton mass $m_{\rm p}$. In {\sc JADE}, this value is computed self-consistently at each iteration by averaging the molecular weights of the individual gas species, weighted by their fractional abundance.

We derive the initial surface density of the dust assuming a constant dust-to-gas ratio of $0.01$, equal to the canonical value for the ISM \citep{bohlin1978}. The initial gas and dust profiles are shown as part of Fig.~\ref{fig:surf_dens}. The gas surface density decreases from $400$~\si{\gram\per\square\centi\meter} at 0.4~au to $1.2$~\si{\gram\per\square\centi\meter} at $R_{\rm c}$. Over this range, the number density of gas particles decreases from $10^{14}$~\si{\per\cubic\centi\meter} to $10^{8}$~\si{\per\cubic\centi\meter}.


\subsubsection{Transport of gas and dust}\label{sec:disc_evol}

We adopt the Shakura-Sunyaev $\alpha$-prescription and parametrise the gas viscosity as $\nu = \alpha\,c_{\rm s}\,H_{\rm p}$, where $\alpha$ quantifies the strength of  \citep{Shakura1973}. Observational evidence suggests low turbulence levels with $\alpha\lesssim10^{-3}$ \citep[e.g.][]{Flaherty2015, Flaherty2018, Rosotti2023}. We assume $\alpha$ to be constant and uniform in the disc, adopting a value of $10^{-3}$.

The viscous evolution of an axisymmetric and geometrically thin $\alpha$-disc is described by the following 1D diffusion equation for the time evolution of $\Sigma_{\rm g}(r,t)$ \citep{Lynden-Bell1974}:
\begin{equation}
    \dfrac{\partial\Sigma_{\rm g}}{\partial t}+\dfrac{1}{r}\dfrac{\partial}{\partial r}(r\,\Sigma_{\rm g}\,v_{\rm g}) = S_{\rm g},
    \label{eq:viscous_gas}
\end{equation}
where $S_{\rm g}$ is the source term that represents a possible external inflow or outflow of gas and  $v_{\rm g}$ is the radial velocity of the gas particles given by the conservation of angular momentum:  
\begin{equation}
    v_{\rm g}=-\dfrac{3\,r^{-1/2}}{\Sigma_{\rm g}}\dfrac{\partial(r^{1/2}\,\Sigma_{\rm g}\,\nu)}{\partial r}. 
    \label{eq:velocity}
\end{equation}
We set $S_{\rm g}=0$, as Class II discs have dispersed their surrounding envelope, and late infall is expected to be negligible in the absence of gravitational interactions with nearby clouds \citep{Gupta2023}.

The viscous accretion timescale is given by $t_{\rm visc}=r^2/\nu$. Considering only the contribution of stellar irradiation to the midplane temperature (see Sec.~\ref{sec:Tmid}), which in our disc model results in a temperature profile $T\propto r^{-1/2}$ with $T=157$~\si{\kelvin} at 1 au, this becomes \citep[see also][]{Booth2019}:
\begin{equation}
    t_{\rm visc}\approx 2.5\times 10^5 \, \left(\dfrac{\alpha}{10^{-3}}\right)^{-1} \, \left(\dfrac{r}{\rm au}\right)\, {\rm yr},
    \label{eq:tvisc}
\end{equation}
corresponding to a viscous timescale of $\sim10$~Myr at the radius $R_{\rm c}$. Equation~\eqref{eq:tvisc} provides an upper limit to the viscous timescale in the inner disc, where viscous dissipation raises the temperature and accelerates the evolution.

We follow \citet{Birnstiel2010} in modelling the evolution of dust grains, taking into account their aerodynamic coupling with the turbulent gas, characterised by their Stokes number $\rm{St}$ \citep{Youdin2007} and assuming that the gas diffusivity is equal to the gas viscosity $\nu$. The surface density profile of the dust component, $\Sigma_{\rm d}(r,t)$, then evolves in time according to the advection-diffusion equation:
\begin{equation}
    \frac{\partial \Sigma_{\rm d}}{\partial t}+\frac{1}{r} \frac{\partial}{\partial r}\left(r \Sigma_{\rm d} v_{\rm d}\right)-\frac{1}{r} \frac{\partial}{\partial r}\left(r\, D_{\rm d}\, \Sigma_{\rm g}\,\frac{\partial}{\partial r}\left(\frac{\Sigma_{\rm d}}{\Sigma_{\rm g}}\right)\right)=S_{\rm d}, 
    \label{eq:drift_dust}
\end{equation}
where $v_{\rm d}$ is the radial velocity of the dust and $D_{\rm d}$ is the dust diffusivity, both of which depend on $\rm{St}$:
\begin{equation}
\begin{array}{ll}
    v_{\rm d}=\dfrac{1}{1+{\rm St}^2}\left(v_{\rm g} + {\rm St}\dfrac{c_s^2}{\Omega_{\rm k}\, r}\,\dfrac{d\,{\rm ln}\,p}{d\,{\rm ln}\,r}\right),\\
    D_{\rm d}=\dfrac{\nu}{1+{\rm St}^2},
\end{array}
\end{equation}
with $p = \rho c_s^2$ being the gas pressure at the midplane. The source term $S_{\rm d}$ in Eq.~\eqref{eq:drift_dust} accounts for dust production and destruction. Since we focus on the radial transport of dust and do not model coagulation or fragmentation processes, we set $S_{\rm d} = 0$ at each time step during the disc evolution. Furthermore, we assume that the dust has already settled in the midplane at the beginning of our simulations (see Sect.~\ref{sec:caveats}).

We numerically solve Eqs.~\eqref{eq:viscous_gas}, \eqref{eq:drift_dust} using the fully implicit integration scheme from \citet{Birnstiel2010}, implemented within {\sc DISKLAB}. The algorithm calculates the flux of gas and dust across the grid cells, tracking the transport of material throughout the disc over time. 

For dust particles, we assume spherical grains with a fixed bulk density, $\rho_{\rm d} = 2.5$~\si{\gram\per\cubic\centi\meter}. This value is physically consistent with the density of CI chondrites or that of a dust-ice mixture with decreased porosity \citep{Consolmagno2008}. We consider three dust grain sizes: 0.1, 20, and 100~\si{\micro\metre}. The smallest size represents submicrometre dust grains that are strongly coupled to the gas, similar to those found in the diffuse ISM \citep{Mathis1977}. At densities higher than those in the ISM, however, dust coagulation favours the formation of much larger grains \citep[e.g.][]{Weidenschilling1993, Dullemond2005, Isella2010, Tazzari2016, Lebreuilly2023, Marchand2023}. The second and third sizes were chosen within the size range ($10-100$~\si{\micro\metre}) for dust grains during the early phases of star and protoplanetary disc formation predicted by \citet{Bate2022}. We consider a single grain size in each simulation to better quantify its impact on the evolution of the disc composition.  

Depending on the local temperature and density conditions in the disc, dust grains exist either in a bare form or coated with ice from the condensation of volatile molecules. As the disc evolves, icy grains undergo local diffusion and mixing. To track the radial redistribution of individual ice species (e.g. H$_2$O, CO$_2$, NH$_3$), we treat each as a separate, independent dust component and assume that they undergo the same dynamical evolution - i.e. we neglect the contribution of the volatile coating to the particle’s Stokes number. The surface density of each ice species is then evolved independently using an advection-diffusion equation analogous to Eq.~\eqref{eq:drift_dust}:
\begin{equation}
    \frac{\partial \Sigma_{{\rm d}, i}}{\partial t}+\frac{1}{r} \frac{\partial}{\partial r}\left(r \Sigma_{{\rm d}, i}\, v_{\rm d}\right)-\frac{1}{r} \frac{\partial}{\partial r}\left(r\, D_{\rm d}\, \Sigma_{\rm g}\,\frac{\partial}{\partial r}\left(\frac{\Sigma_{{\rm d},i}}{\Sigma_{\rm g}}\right)\right)=0, 
    \label{eq:drift_dust_index}
\end{equation}
where $i = 1, \dots, 184$ corresponds to the number of ice species included in our model. Since all species share the same dynamics, summing the solutions of Eq.~\eqref{eq:drift_dust_index} over all $i$ is numerically equivalent to solving Eq.~\eqref{eq:drift_dust} for the total surface density of ice, defined as the sum of contributions from all ice species.

Since the diffusion coefficient approaches the turbulent gas diffusivity in the limit ${\rm St} \rightarrow 0$, Eq.~\eqref{eq:drift_dust} also applies to the advection and diffusion of gas particles. Following the same approach as for icy grains, we track the radial transport of individual volatile species in the gas phase by solving equations analogous to Eq.~\eqref{eq:drift_dust_index}, with the index $i$ running over the 484 gas-phase species included in our model. For gas particles, we adopt a nominal size of $10^{-10}$~cm to ensure tight coupling.

To better interpret the temporal changes in the chemical composition of the disc due to mass transport, it is useful to estimate the radial drift timescale, $t_{\rm drift}=r/v_{\rm d}$, for a dust grains of size $a$. We adopt the parametrisation from \citet{Booth2019}:
\begin{equation}
    t_{\rm drift}\approx 1\times 10^8 \, \left(\dfrac{\Sigma_0}{10^{2}\,{\rm g\,cm^{-2}}}\right)\, \left(\dfrac{a}{\rm \mu m}\right)^{-1}\, {\rm yr}.
    \label{eq:tdrift}
\end{equation}
This expression is rigorously valid for a density structure where $\Sigma_{\rm g} \propto r^{-1}$, slightly steeper than the one adopted in this work. However, it still provides a useful approximation for a qualitative analysis of the numerical solution (Sect.~\ref{sec:res_volatiles}).

\subsubsection{Midplane temperature}\label{sec:Tmid}

\begin{figure*}
    \includegraphics[width = \textwidth]{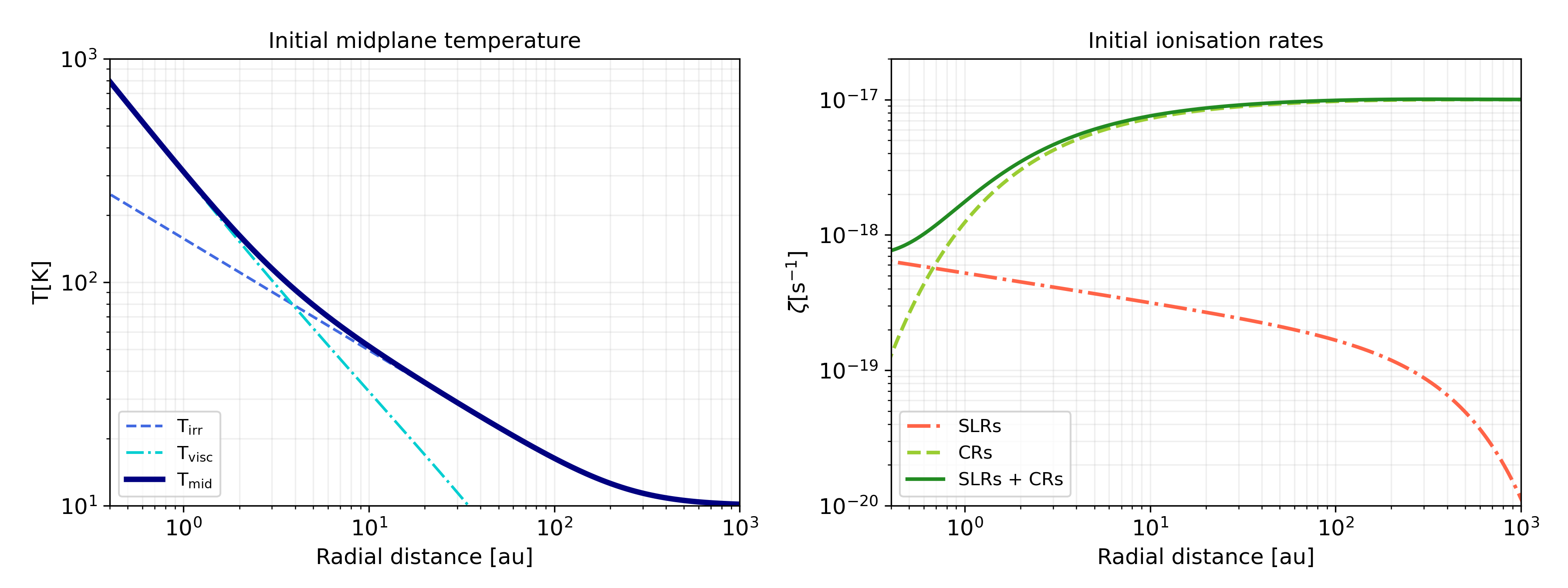}
    \caption{\textit{Left:} Initial radial profile of the midplane temperature (solid) together with the contributions of stellar irradiation (dashed) and viscous heating (dash-dotted). The temperature reaches a minimum value of 10~\si{\kelvin} in the outer disc. \textit{Right:} Initial ionisation profiles in the midplane. The orange (dash-dotted) curve indicates the contribution of SLRs and characterises the ionisation environment in the \emph{low} scenario. The green (solid) curve shows the total ionisation taking into account an additional contribution from CRs (light green curve) and characterises the ionisation environment in the \emph{high} scenario.} 
    \label{fig:tmid_ion}
\end{figure*}

To account for the feedback of disc evolution on the midplane temperature, we consider two primary heating mechanisms: the reprocessing of stellar irradiation and the internal energy dissipation from viscous accretion \citep{Dalessio2005}. Together, these mechanisms contribute to the midplane temperature structure as follows:
\begin{equation}
    T_{\rm mid} = (\,T_{\rm irr}^4 + T_{\rm visc}^4\,)^{1/4}, 
    \label{eq:tmid}
\end{equation}
where $T_{\rm irr}$ and $T_{\rm visc}$ represent the contributions from stellar irradiation and viscous heating, respectively. To prevent the temperature in the outer disc from dropping to unphysically low values, we set a lower limit of 10~\si{\kelvin} wherever $T_{\rm mid}$ would otherwise fall below this threshold. This is consistent with the lowest midplane temperatures inferred for the disc around HD 163296 from CO line emission \citep{dullemond2020}.

The irradiation temperature is calculated by balancing the heating from stellar irradiation with the radiative cooling of the disc, giving: 
\begin{equation}
    T_{\rm irr}^4 = \dfrac{1}{2}\, \dfrac{\sin{\alpha(r)}}{\sigma_{\rm B}}\, \dfrac{L_\star}{4\pi r^2},  
    \label{eq:tirr}
\end{equation}
where $\alpha(r)$ is the grazing angle between the incident stellar rays and the disc surface, $\sigma_{\rm B}$ is the Stefan–Boltzmann constant, and $L_\star$ is the stellar luminosity \citep{Chiang1997, Dullemond2001}. For our geometrically thin, non-self-shadowed disc, we assume that stellar light reaches the disc surface at a constant, shallow grazing angle $\alpha \ll 1$, and adopt a global value of $\sin{\alpha}\approx \alpha = 0.05$.

The contribution of viscous heating to the midplane temperature is estimated following the approach of \citet{Nakamoto1994}:
\begin{equation}
    T_{\rm visc}^4 = \dfrac{9}{8\sigma_{\rm B}}\,\left(\dfrac{\tau_{\rm Ross}/2+1}{1-e^{-2\,\tau_{\rm Ross}}}\right)\,\Sigma_{\rm g}(r)\,\nu(r)\,\Omega_{\rm k}(r)^2, 
    \label{eq:tempvisc}
\end{equation}
where $\tau_{\rm Ross}=\Sigma_d\,\kappa_{d,\rm Ross}$ is the Rosseland mean optical depth, and $\kappa_{d,\rm Ross}$ is the Rosseland mean opacity of the dust. Solutions for $T_{\rm visc}$ are calculated iteratively using Brent’s root-finding algorithm, assuming a constant dust opacity of $\kappa_{d,\rm Ross} = 10^3$ throughout the disc and for all simulated grain sizes. 

From $T_{\rm mid}$, the sound speed is automatically calculated at each time step during the disc evolution and is given by: 
\begin{equation}
    c_s^2=\dfrac{k_{\rm B}\,T_{\rm mid}}{\mu_{\rm g}\,m_p},
\end{equation}
where $k_{\rm B}$ is the Boltzmann's constant. The initial midplane temperature profile is shown in Fig.~\ref{fig:tmid_ion}. In the inner disc, $T_{\rm visc}$ dominates due to higher gas densities, whereas $T_{\rm irr}$ prevails in the outer disc. We assume that the stellar luminosity remains constant at the present-day solar value over the timescale of our simulations, so that $T_{\rm irr}$ varies only with the distance from the star. In contrast, $T_{\rm visc}$ evolves over time due to changes in the surface density.

\subsubsection{Ionisation environment}

We model the ionisation environment of the disc midplane following the same approach as in \citet{Eistrup2016} and \citet{Pacetti2022}. We consider two scenarios with low and high ionisation levels. 

In the \emph{low} scenario, the only source of ionising radiation in the midplane is the decay of short-lived radionuclides (SLRs). The associated ionisation rate per H$_2$ molecule is parametrised as follows \citep{Cleeves2013b}:
\begin{equation}
	\zeta_\text{SLR}(r)=1.25\times 10^{-19}\,\left(\dfrac{\Sigma_{\rm g}(r)}{\text{g}\,\text{cm}^{-2}}\right)^{0.27}\, \text{s}^{-1},
 \label{eq:slrs_p2}
\end{equation}

In the \emph{high} scenario, ionisation from SLRs is enhanced by an additional contribution from external cosmic rays (CRs), taken as the standard cosmic-ray ionisation rate, $\zeta^{\rm H_2}\approx 10^{-17}$ \citep{Spitzer1968}, attenuated by the gas surface density \citep{Umebayashi1981,Eistrup2016}: 
\begin{equation}
	\zeta_\text{CR}(r)=10^{-17}\,\exp\left(\dfrac{-\Sigma_{\rm g}(r)}{96\,\text{g}\,\text{cm}^{-2}}\right)\, \text{s}^{-1}. 
 \label{eq:crs_p2}
\end{equation}
The initial ionisation rates for the two scenarios are shown in Fig.~\ref{fig:tmid_ion}. 


\subsection{Disc chemical model}\label{sec:chem_model}
The compositional model of the disc builds upon the one presented by \citet{Pacetti2022}, with updated reference abundances and a revised parametrisation of the radial abundance profile of refractory organic carbon.

\subsubsection{Setting the initial conditions for chemistry}\label{sec:comp_model}

The disc model assumes that the planet-forming material is distributed across three reservoirs: rocks, volatiles (gas and ices), and refractory organic carbon. We determine the initial elemental abundances in these reservoirs starting from a protosolar gas mixture, following the approach of \citet{Pacetti2022}. 

The rocky component hosts all elements except the noble gases and includes a fraction of the total available H, C, O, and N, with the remainder of these elements considered to reside in volatile form (see below). We assume that the rock-forming elements were originally present in meteoritic proportions in the protosolar mixture, adopting the meteoritic abundances of CI chondrites from \citet{Lodders2010}. We focus on the elemental budget of the rocky component and do not model its specific mineralogy (e.g silicates, carbonaceous compounds, or other refractory phases). Furthermore, following \citet{Palme2014} and \citet{Turrini2023}, we assume a fixed composition for the rocky material and do not model the sublimation of its constituents at temperatures exceeding 300~\si{\kelvin} within the innermost 1~au \citep{Fegley2010}.

For H, the noble gases He, Ne, and Ar, and the volatile fractions of C, O, and N, the initial abundances are estimated from the updated solar photospheric abundances of \citet{Asplund2021}. We apply a correction factor to account for atomic diffusion, which causes elements heavier than H to partially settle out of the Sun’s outer convective zone into its radiative interior. Following \citet{Vinyoles2017} and \citet{Asplund2021}, we adopt a correction factor of 0.07 dex for He. For C, O, N, Ne, and Ar, we use a correction factor of 0.03 dex, as predicted by the recent model of \citet{Eggenberger2022}, which includes internal rotation and better reproduces Li depletion in the Sun and helioseismological constraints.

The protosolar and meteoritic abundances describing the original mixture are listed in Tab.~\ref{tab:proto_abund}, along with the abundances in the residual gas after condensation of the rock-forming elements. All abundances in this study are given with respect to the total number of H nuclei.

\begin{table*}
    \caption{Initial conditions for chemistry: Elemental abundances with respect to total H nuclei in the initial gas mixture (protosolar), in rocks (meteoritic) and in the residual gas after condensation of rocks.}\label{tab:proto_abund}
\begin{center}
\begin{tabular}{ccccccccc}
    \hline\hline
    \noalign{\smallskip}
    \textbf{Element} & \textbf{Protosolar} & \textbf{Meteoritic} & \textbf{Residual gas} & \textbf{Element} & \textbf{Protosolar} & \textbf{Meteoritic} & \textbf{Residual gas}\\
    \noalign{\smallskip}
    \hline 
    \noalign{\smallskip}
    H & $1.0$ & $1.7\times10^{-4}$ & $1.0$ & Ca & $2.3\times10^{-6}$ & $2.3\times10^{-6}$ & - \\
    He & $9.5\times10^{-2}$ & $<10^{-10}$ & $9.5\times10^{-2}$ & Na & $1.8\times10^{-6}$ & $1.8\times10^{-6}$ & - \\
    O & $5.2\times10^{-4}$ & $2.6\times10^{-4}$ & $2.6\times10^{-4}$ & Ni & $1.7\times10^{-6}$ & $1.7\times10^{-6}$ & - \\
    C & $3.1\times10^{-4}$ & $2.6\times10^{-5}$ & $2.8\times10^{-4}$ & Cr & $4.6\times10^{-7}$ & $4.6\times10^{-7}$ & - \\
    Ne & $1.2\times10^{-4}$ & $<10^{-10}$ & $1.2\times10^{-4}$ & Cl & $3.5\times10^{-7}$ & $3.5\times10^{-7}$ & - \\
    N  & $7.2\times10^{-5}$ & $1.9\times10^{-6}$ & $7.1\times10^{-5}$ & Mn & $2.9\times10^{-7}$ & $2.9\times10^{-7}$ & - \\ 
    Mg & $4.3\times10^{-5}$ & $4.3\times10^{-5}$ & - & P & $2.8\times10^{-7}$ & $2.8\times10^{-7}$ & - \\
    Si & $3.5\times10^{-5}$ & $3.5\times10^{-5}$ & - & K & $1.2\times10^{-7}$ & $1.2\times10^{-7}$ & - \\
    Fe & $3.2\times10^{-5}$ & $3.2\times10^{-5}$ & - & Ti & $9.3\times10^{-8}$ & $9.3\times10^{-8}$ & - \\
    S & $1.4\times10^{-5}$ & $1.4\times10^{-5}$ & - & F & $2.8\times10^{-8}$ & $2.8\times10^{-8}$ & - \\
    Al & $3.0\times10^{-6}$ & $3.0\times10^{-6}$ & - & V & $8.5\times10^{-9}$ & $8.5\times10^{-9}$ & - \\
    Ar & $2.6\times10^{-6}$ & $<10^{-10}$ & $2.6\times10^{-6}$ & & &\\
    \noalign{\smallskip}
    \hline
\end{tabular}
\end{center}
\tablefoot{Protosolar abundances of volatiles and noble gases from \citet{Asplund2021} and corrected for sinking \citep{Vinyoles2017, Eggenberger2022}. Meteoritic abundances from \citet{Lodders2010}.}
\end{table*}

Based on \citet{Lee2010}, \citet{Gail2017}, \citet{vantHoff2020}, \citet{Li2021}, and \citet{Nazari2023}, we introduce an additional chemical species, $C_{\rm ref}$, to trace the evolution of refractory organic carbon. Similar to rocks, refractory organic carbon is treated as a reservoir for semi-refractory organic carbon, which remains inert throughout most of the disc and undergoes chemical processing only within its sublimation radius, where its volatile component returns to the gas phase. Following \citet{Mordasini2016} and \citet{Cridland2019a}, we model carbon erosion in the disc with a power-law abundance profile for $C_{\rm ref}$ between 1 and 5~au:
\begin{equation}\label{eq:cref}
    C_{\rm ref}(r) = \left\{
    \begin{array}{lll}
        C_{\rm ref,\, \rm min}            & \quad r \leq 1\,\rm au \\
        C_{\rm ref,\, \rm min}\cdot r^c   & \quad 1 \,\rm au < r < 5\,\rm au \\
        C_{\rm ref,\, \rm max}            & \quad r \geq 5\,\rm au, 
    \end{array}
    \right.  
\end{equation}
where $C_{\rm ref,\, \rm min}$ is set to $10^{-20}$ and $C_{\rm ref,\, \rm max}$ to $1.9 \times 10^{-4}$, which corresponds to about $60\%$ of the total carbon in the original protosolar mixture (see Tab.~\ref{tab:proto_abund}), consistent with the carbon content observed in solar-system comets \citep{Bardyn2017}. These boundary conditions yield a power-law index of $c \sim 23.3$. We treat $C_{\rm ref}$ as a semi-refractory, carbon-bearing species transported by dust grains, with its initial abundance set by the profile in Eq.~\eqref{eq:cref}. Within 5~au, we assume that $C_{\rm ref}$ is released back into the gas phase as atomic carbon, following a complementary profile to that in Eq.~\eqref{eq:cref}. To simulate the advection and diffusion of $C_{\rm ref}$ during disc evolution, we update the abundance of atomic C in the gas phase at each time step by adding the fraction of $C_{\rm ref}$ that has crossed 5~au and sublimated. 

Overall, the refractory component of our disc contains approximately $50\%$ of the protosolar O, $\sim 68\%$ of C (with $8\%$ in rocks and $60\%$ in $C_{\rm ref}$), and $\sim 3\%$ of N. The volatile component accounts for the remaining $50\%$ of O, $32\%$ of C, and $97\%$ of N. Table \ref{tab:proto_abund} lists the total abundances of elements available for forming volatiles in the residual gas after the condensation of rocks. For C, this abundance decreases to approximately $9.1\times10^{-5}$ beyond 5~au, where $C_{\rm ref}$ remains in solid form. 

We simulate the chemical evolution of the disc using two different sets of initial abundances for the volatile component, listed in Tab.~\ref{tab:init_abund} along with the adopted binding energies, E$\rm_B$(K). These two sets correspond to the \emph{inheritance} and \emph{reset} scenarios described in \citet{Eistrup2016,Eistrup2018} and \citet{Pacetti2022}. In the inheritance scenario, C, O, and N are initially locked in seven key volatile molecules, representing a disc that has inherited its composition from the prestellar phase. We partition these elements among their seven major molecular carriers using the same molecular ratios as in \citet{Eistrup2016}, which are representative of interstellar ice \citep{Marboeuf2014, Boogert2015}. In the reset scenario, by contrast, C, O, and N are initialised in their atomic form, representing an extreme case in which all molecules have been fully dissociated due to strong heating by the protostar. In both the inheritance and reset scenarios, the atomic carbon abundance reported in Tab.~\ref{tab:init_abund} applies to disc regions beyond 5~au. Inside 5~au, the C abundance is enhanced by the contribution of sublimated carbon from $C_{\rm ref}$, as described above.

\begin{table*}
    \caption{Initial abundances of volatile species with respect to total H-nuclei in the inheritance and reset scenarios, together with their binding energies.}\label{tab:init_abund}
\begin{center}
\begin{tabular}{ccccccccc}
    \hline\hline
    \noalign{\smallskip}
    \textbf{Species} & \textbf{Inheritance} & \textbf{Reset} & \textbf{E$_{\rm B}$(K)} & & \textbf{Species} & \textbf{Inheritance} & \textbf{Reset} & \textbf{E$_{\rm B}$(K)} \\
    \noalign{\smallskip}
    \hline 
    \noalign{\smallskip}
    H & $5.0\times10^{-5}$ & $9.1\times10^{-5}$ & 600 &  & H$_2$O & $1.5\times10^{-4}$ & - & 5770 \\
    H$_2$ & $5.0\times10^{-1}$ & $5.0\times10^{-1}$ & 430 &  & CO & $3.0\times10^{-5}$ & - & 855 \\
    He & $9.5\times10^{-2}$ & $9.5\times10^{-2}$ & 100 &  & CO$_2$ & $3.0\times10^{-5}$ & - & 2990 \\
    C & - & $9.1\times10^{-5}$ & 800 &  & CH$_4$ & $9.0\times10^{-6}$ & - & 1090 \\
    O & - & $2.6\times10^{-4}$ & 800 &  & CH$_3$OH & $2.2\times10^{-5}$ & - & 4930 \\
    N & - & $7.1\times10^{-5}$ & 800 &  & N$_2$ & $2.4\times10^{-5}$ & - & 790 \\
    & & & & & NH$_3$ & $2.4\times10^{-5}$ & - & 3130 \\
    \noalign{\smallskip}
    \hline
\end{tabular}
\end{center}
\tablefoot{Binding energies from \citet{Eistrup2016}. Note that the total elemental abundances listed here match those of the residual gas in Tab.~\ref{tab:proto_abund}, except for carbon, which excludes the fraction locked in $C_{\rm ref}$ beyond 5~au.}
\end{table*}

\subsubsection{Chemical evolution of volatile molecules}\label{sec:chem_evol}

The evolution of the abundances of volatile molecules in the disc midplane is calculated using the two-phase chemical kinetics code by \citet{Walsh2015}. The chemical network used in this study consists of 668 species (gas + ice) participating in 8385 chemical reactions, which include gas-phase reactions, gas-grain interactions, and grain-surface chemistry. This corresponds to the \emph{full chemical network} used in \citet{Eistrup2016, Eistrup2018} and on which the disc chemical model of \citet{Pacetti2022} is also based. A detailed description of the chemical network can be found in \citet{Walsh2015} and \citet{Eistrup2016}.  

The high opacity of protoplanetary discs prevents stellar UV photons from penetrating deeper than the upper layers of the disc atmosphere. As a result, photochemistry in the midplane is strongly inhibited. To simulate this condition, we assume a fixed extinction coefficient $A_v=10$ throughout the disc grid. UV photons can still be generated internally via the interaction of CRs with H$_2$ molecules \citep{Eistrup2016} and are responsible for photodesorption and CR-induced photoreactions. Following \citet{Eistrup2016, Eistrup2018}, we neglect the contribution of X-rays and consider SLRs and CRs as the main ionisation sources in the midplane.

Dust grains actively participate in the chemical evolution by providing surfaces for adsorption and desorption processes. We assume spherical grains with fixed radius and density (see Sect.~\ref{sec:disc_evol}), and use the radial dust surface density profile derived from the physical model to compute the abundance of grains relative to the total number of H nuclei at each location in the disc. This abundance is updated at each time step during the disc evolution. The grain size also determines the number of available sites for surface reactions, which in turn affects the efficiency of gas-grain interactions and surface chemistry. Large grains can host more reaction sites on their surface than small grains. However, due to their larger size, they are comparatively less abundant, which ultimately leads to less efficient grain chemistry \citep[e.g.][]{Navarro2024}.

We assume thermal equilibrium between the gas and dust, which is expected in the disc midplane due to the high densities. Additionally, we neglect the thermodynamic effects of the chemical reactions on the gas and dust temperatures in order to limit the computational cost of the simulations.


\subsection{Population of planetesimals}\label{sec:plts}

We assume that $50\%$ of the dust is converted into km-sized planetesimals uniformly throughout the disc by $10^5$~yr. Once planetesimals form, they no longer participate in chemical interactions with the gas or the remaining $50\%$ of dust in the disc. Their large size results in a very low total surface area per unit volume, making surface chemistry extremely inefficient.

The presence of planetesimals also reduces the amount of ice material that can be released into the gas phase due to thermal processing \citep{Gkotsinas2024}. Planetesimals are therefore treated as a sink for refractory material and ice, which is subtracted from the initial chemical budget and does not participate in the subsequent chemical evolution of the disc. Their ice content is determined by the composition of the ice phase of the disc at their formation location at $10^5$~yr.


\section{Chemistry in a static disc}\label{sec:res_static_model}

\begin{figure*}[ht]
\centering
\includegraphics[width = 0.8 \textwidth]{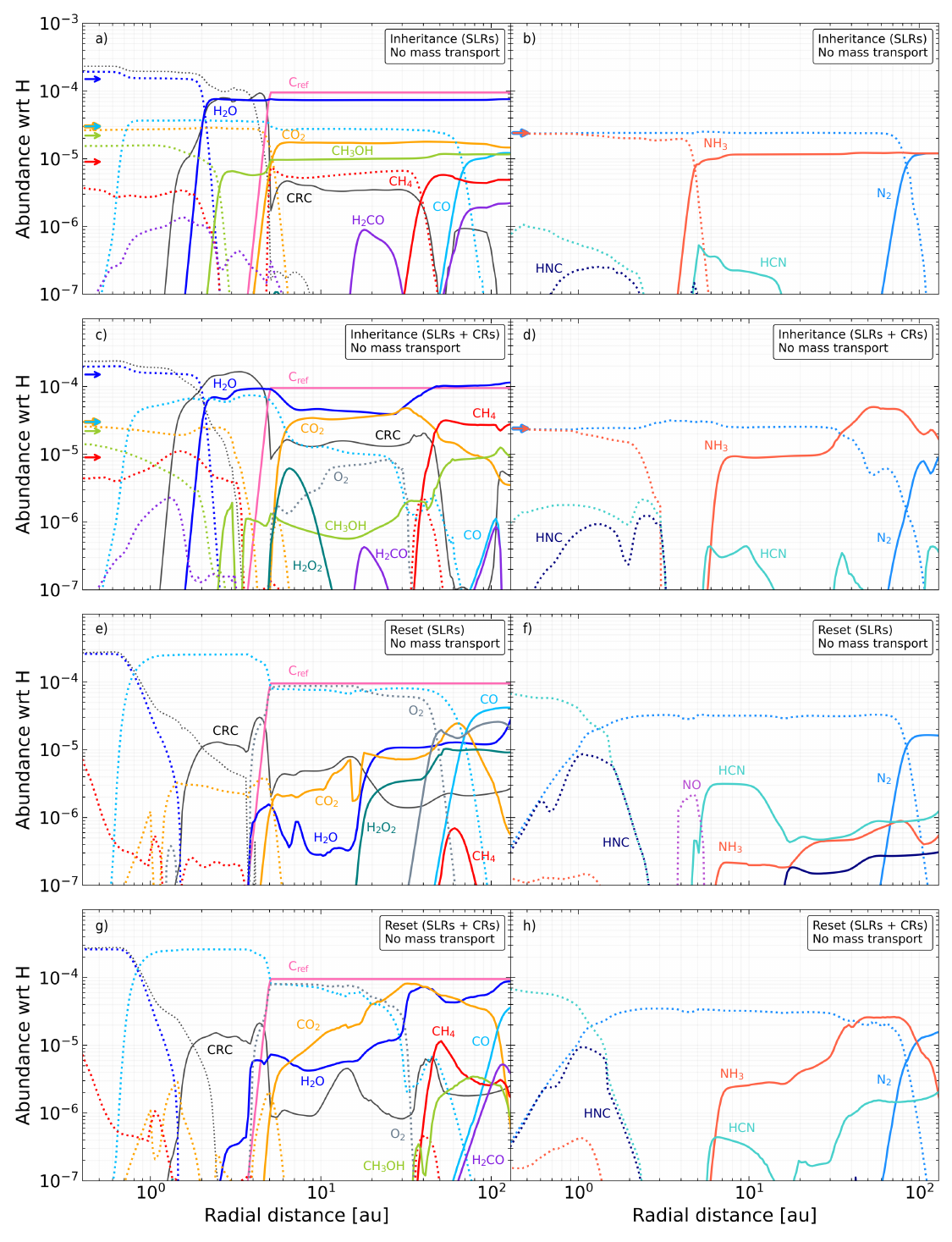}
\caption{Radial abundance profiles of the main C-, O- (left), and N-bearing (right) volatile molecules in the gas phase (dotted lines) and in the ice phase (solid lines) in the disc midplane after 3~Myr of chemical evolution without mass transport in the disc (stationary disc). The results are shown for the four chemical scenarios assuming a grain size of $0.1$~\si{\micro\meter}. The horizontal small arrows indicate the initial abundances of H$_2$O, CO, CO$_2$, CH$_3$OH, CH$_4$, N$_2$ and NH$_3$ in the gas phase in the inheritance scenario (see also Tab. \ref{tab:init_abund}); note that CO and CO$_2$, as well as N$_2$ and NH$_3$, share the same initial abundance. The black profile (CRC), represents the cumulative abundance of additional carbon-rich compounds, including carbon chains, hydrocarbons, nitriles, and other complex organics that are not among the dominant volatile carriers.} 
\label{fig:static_disc}
\end{figure*}

To quantify the relative effects of the physical and chemical evolution on the distribution of volatiles in the midplane, we first simulate the chemical evolution of a stationary disc populated by grains of size $0.1$~\si{\micro\meter}. We perform simulations for the two different sets of initial chemical abundances (inheritance and reset) and ionisation rates (SLRs and SLRs+CRs). The surface density and temperature of the gas and the dust are set by Eqs.~\eqref{eq:sigma_gas_paperII}, \eqref{eq:tmid}. Figure \ref{fig:static_disc} shows the resulting abundance profiles of the main molecular carriers of C, O, and N at 3~Myr. 

Globally, our results for the static disc are consistent with the trends found by \citet{Eistrup2016} at 1~Myr, which we used to benchmark our implementation of chemical kinetics in {\sc JADE}. The main differences arise from the disc structure and the initial gas-phase metallicity, which in our case reflects a more detailed compositional model that includes refractory and semi-refractory components. 

In the inheritance-low scenario (panels \emph{a} and \emph{b}), chemical evolution largely preserves the initial molecular budget. Thermal desorption and freeze-out dominate, driving the phase transitions at the snowlines. The difference between the abundances of volatiles in the ice phase (solid lines) and the initial molecular abundances (horizontal dash-dotted lines) stems from our assumption that $50\%$ of the dust is converted into chemically inert planetesimals early in the disc's evolution and uniformly throughout the disc, effectively halving the abundances of volatiles in the remaining ice phase.

Increases in the ionisation rate favour direct ionisation reactions and photoreactions by CR-induced UV photons. In the inheritance-high scenario (panels \emph{c} and \emph{d}), the CH$_4$ gas is depleted between its snowline ($\sim40$~au) and that of CO$_2$ ($\sim5$~au) due to its CR-driven conversion into CO$_2$. We find that this depletion persists even after 3~Myr, consistent with the results of \citet{Eistrup2018}. Similarly, CRs drive the destruction of NH$_3$ in favour of N$_2$ between 3 and 6~au, that is, at temperatures ranging from 120 to 70~\si{\kelvin} and number densities from $5\times 10^{12}$ to $10^{12}$~\si{\per\cubic\centi\meter}. Our results also indicate that NH$_3$ ice becomes the primary nitrogen reservoir in the outer disc under high ionisation conditions and its abundance doubles compared to the initial value near the N$_2$ snowline ($\sim85$~au). The inclusion of surface chemistry allows us to capture additional effects, such as CR-induced photodissociations of ice species in situ, which are responsible, for example, for the depletion of CH$_3$OH and CO beyond their snowlines ($\sim2.5$~au and $70$~au, respectively).

For both ionisation levels, the sublimation of $C_{\rm ref}$ within 5~au (see Sect.~\ref{sec:comp_model}) favours the formation of carbon-rich radicals and chains, hydrocarbons (e.g C$_2$H$_2$ and C$_2$H$_6$), nitriles, and promotes the conversion of small organic molecules (e.g. CH$_4$ and CH$_3$OH) into larger organics and complex organic molecules (e.g. CH$_3$CHO, CH$_3$COOH, CH$_3$COCH$_3$). The cumulative abundance of these species - here collectively referred to as "carbon-rich compounds" (CRC) for simplicity - is represented by the black profile in Fig.~\ref{fig:static_disc}, and reaches levels comparable to H$_2$O within the $C_{\rm ref}$ snowline (panels \emph{a} and \emph{c}). This chemical pathway is particularly efficient when $C_{\rm ref}$ is released as atomic carbon into the gas phase. The effects of releasing $C_{\rm ref}$ in other chemical forms (e.g. hydrocarbons) or at higher temperatures \citep[e.g.][]{Gail2017, vantHoff2020} are not explored here and will be addressed in future work. 

When considering the reset scenarios (panels \emph{e} through \emph{h}), chemistry is overall more active. For both ionisation levels, the main gas-phase carriers of C, O, and N are H$_2$O, CO, O$_2$, HCN, and N$_2$, while CO$_2$, CH$_4$, and NH$_3$ are significantly depleted. Specifically, H$_2$O and HCN dominate within 1~au, while CO and N$_2$ become the primary carriers from about 1~au to their respective snowlines beyond 60~au. CO, in particular, forms readily in these scenarios from available atomic C and O, especially within the $C_{\rm ref}$ snowline, where its abundance increases by a factor of 2 compared to the outer disc (panels \emph{e} and \emph{g}). Between 4 and 40~au, O$_2$ rises to the same level as CO, reaching an abundance of about $8\times 10^{-5}$, while CH$_4$ remains below $10^{-7}$. Beyond 40~au, O$_2$ condenses as ice, but only under low ionisation conditions. In the reset-high scenario, O$_2$ is efficiently destroyed in the outer disc due to its high reactivity to cosmic ray-induced photoreactions \citep[see also][]{Eistrup2016}. Another difference between the low and high ionisation scenarios is in the ice-phase abundances of H$_2$O, CO$_2$, and NH$_3$, which are all systematically higher in the reset-high scenario. In contrast to the inheritance scenarios, the abundance of CRC is significantly reduced throughout the disc due to the efficient and early conversion of atomic carbon into CO and the limited availability of molecular precursors for complex organic synthesis.

Overall, variations in the initial chemical conditions mainly affect the oxygen and nitrogen chemistry. Specifically, the gas phase in a disc midplane that has undergone a chemical reset is more enriched in oxygen compared to the inheritance scenario and undergoes significant depletion of NH$_3$ in favour of other N-bearing species, such as HCN. This emphasises an intrinsic difference between the inheritance and reset chemical scenarios, which directly impacts the midplane elemental ratios (see Sect.~\ref{sec:res_ratios}). 

\section{Transport of gas and dust}\label{sec:res_transport}

\begin{figure*}
\centering
\includegraphics[width = \textwidth]{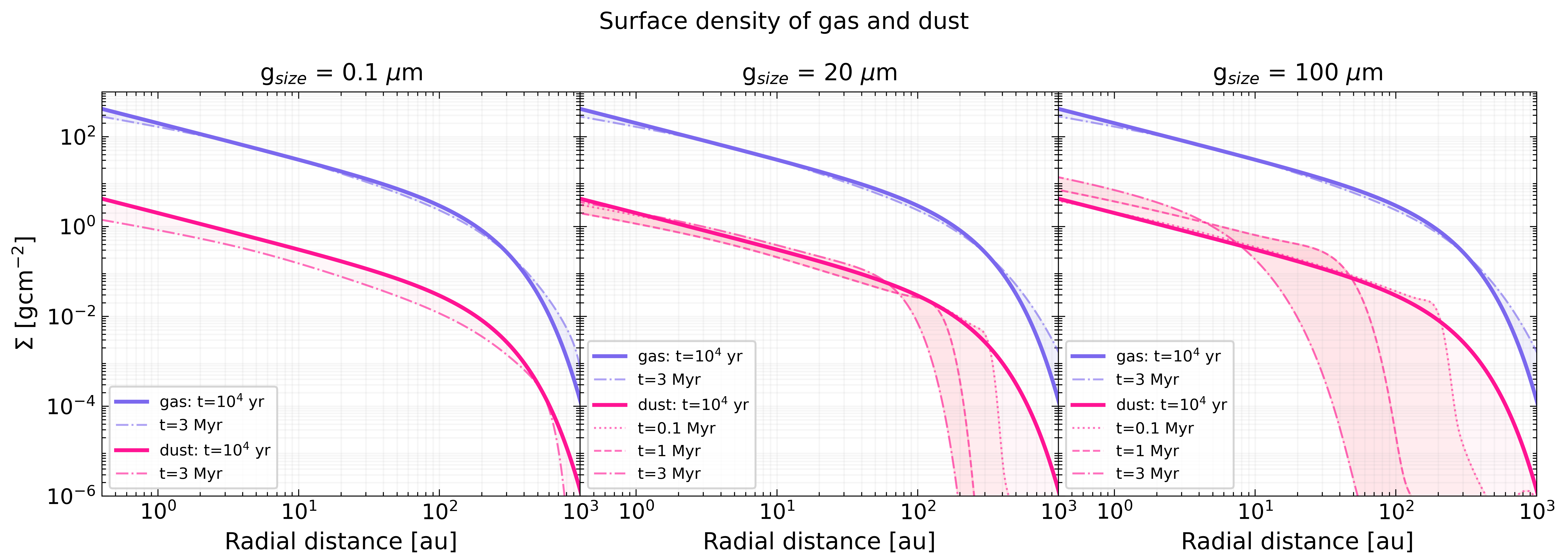}
\caption{Evolution of the surface density of the gas (violet curves) and the dust (pink curves) in the three scenarios with grains of size 0.1~\si{\micro\meter}, 20~\si{\micro\meter}, and 100~\si{\micro\meter}, assuming $\alpha=10^{-3}$. The solid lines represent the initial conditions at $t=10^4$~yr, while the dotted, dashed, and dash-dotted lines show the evolved profiles at 0.1, 1, and 3~Myr, respectively. The effect of the inward radial drift of the dust relative to the gas becomes stronger with increasing grain size. Note that the surface density of the dust does not include the mass contribution of the ice mantles beyond the snowlines.} 
\label{fig:surf_dens}
\end{figure*}

The results for the stationary disc isolate the effects of chemical processes on the composition of the midplane, providing a basis for interpreting the chemical outcomes of our fully integrated model. Before delving into the chemistry, we first examine the effects of mass transport on the radial distribution of gas and dust in the midplane, which in turn governs the radial redistribution of volatiles in the disc.

Figure~\ref{fig:surf_dens} compares the evolution of the surface density of gas (violet) and dust (pink) in the three grain size scenarios. The dust surface density represents the mass distribution of dust grains, excluding the contribution from ice mantles. While our model accounts for ice formation beyond snowlines, we neglect its impact on grain dynamics, and therefore do not include the mass of the ice mantles in the plotted dust surface densities. 

In line with the predictions of viscous evolution, the majority of the gas in the inner disc is transported inwards and ultimately lost to the star, while a smaller fraction is moved outwards to larger radii (see the dash-dotted violet curve).

\begin{figure}
\centering
\includegraphics[width = \columnwidth]{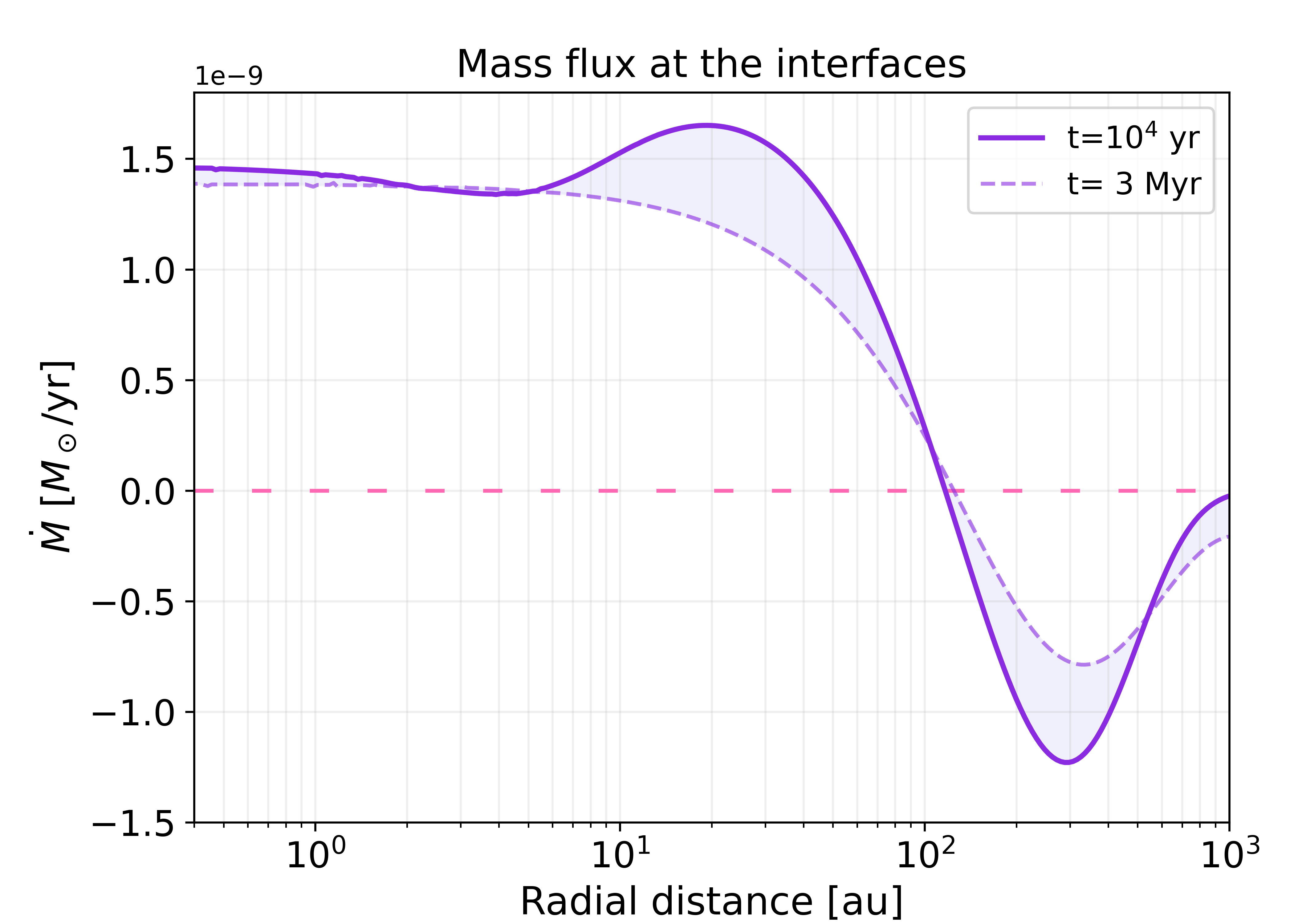}
\caption{Radial profiles of the mass accretion flux in the disc midplane in units of solar masses of gas per year. The plot shows the mass flux computed at the edges of each grid point at $t=10^4$~yr (solid curve) and $t=3$~Myr (dashed curve), corresponding to the beginning and the end of our simulations, respectively. The accretion flux is positive up to about 120~au and then becomes negative. A positive (negative) accretion flux indicates gas transport towards (away from) the star, which is consistent with the predictions of viscous evolution.} 
\label{fig:acc_rate}
\end{figure}

Over the first 3~Myr, the mass accretion flux, $\dot{M}$, evolves as shown in Fig.~\ref{fig:acc_rate}. $\dot{M}$ represents the mass flux calculated at the interface between adjacent grid points. At the disc's inner edge, the accretion rate onto the star is approximately $1.4 \times 10^{-9} M_\odot/{\rm yr}$. Over the timescale of our simulations, this results in the accretion of about $9\%$ of the initial gas mass. Specifically, the disc's gas mass decreases from approximately $5.4\times10^{-2}M_\odot$ to $4.9\times10^{-2}M_\odot$ over 3~Myr. Notably, about $1\%$ of the disc mass is initially located within 4~au of the star and is accreted within the first $3.5 \times 10^{5}$~yr. As a result, the inner disc midplane is depleted in less than 1~Myr and is continuously replenished with material from the outer disc. If volatiles (gas and ice) are less abundant in the outer disc than in the inner disc - as in our model, where part of the initial ice budget is sequestered by planetesimals that chemically decouple from the gas - the mass redistribution caused by viscous accretion alters the overall elemental balance of volatiles in the inner disc. This effect, independent of chemical processes but captured by our model, is further discussed in Sect.~\ref{sec:res_volatile_strong}.

The dynamical evolution of dust grains is strongly influenced by the strength of the gas-dust coupling. As long as $\rm{St}\ll1$, the grains remain tightly coupled to the gas, and the radial drift acts on much longer timescales than the viscous timescale of the disc ($t_{\rm drift}\gg t_{\rm visc}$). This scenario applies to submicrometre-sized grains (left panel in Fig.~\ref{fig:surf_dens}). Dragged along by the gas, these grains accrete onto the star, resulting in a 3-fold decrease in dust surface density within 100~au after 3~Myr.

As the grain size increases and the Stokes number approaches 1, the grains gradually decouple from the gas and experience a stronger inward drift. In the scenario with 20~\si{\micro\meter} grains (central panel in Fig.~\ref{fig:surf_dens}), $t_{\rm drift}$ becomes shorter than $t_{\rm visc}$ from about 50~au outwards. The dust in this region drifts inwards on a timescale of about 10~Myr, partially offsetting the initial dust mass loss within 50~au due to gas drag. After 3~Myr, all grains are located within 200~au, with most of the mass concentrated within 100~au, and the dust surface density within 50~au has returned to approximately its initial value.

For larger grains, radial drift dominates over viscous evolution and significantly influences the radial extent of the dust disc. In the scenario with 100~\si{\micro\meter}-sized grains (right panel in Fig.~\ref{fig:surf_dens}), the disc region where the dust drifts faster than the gas extends from about 5-10~au outward. Here, radial drift operates on a timescale of about 2~Myr, leaving only a small amount of dust beyond 50~au ($\Sigma_{\rm d} \lesssim 10^{-6}$~\si{\gram\per\square\centi\meter}) by the end of the simulation. Meanwhile, the dust mass within the first 5~au increases by a factor of up to $\sim 3$, as does the dust-to-gas mass ratio.

\section{Chemistry in an evolving disc}\label{sec:res_volatiles}

\begin{figure*}
\centering
\includegraphics[width = 0.9\textwidth]{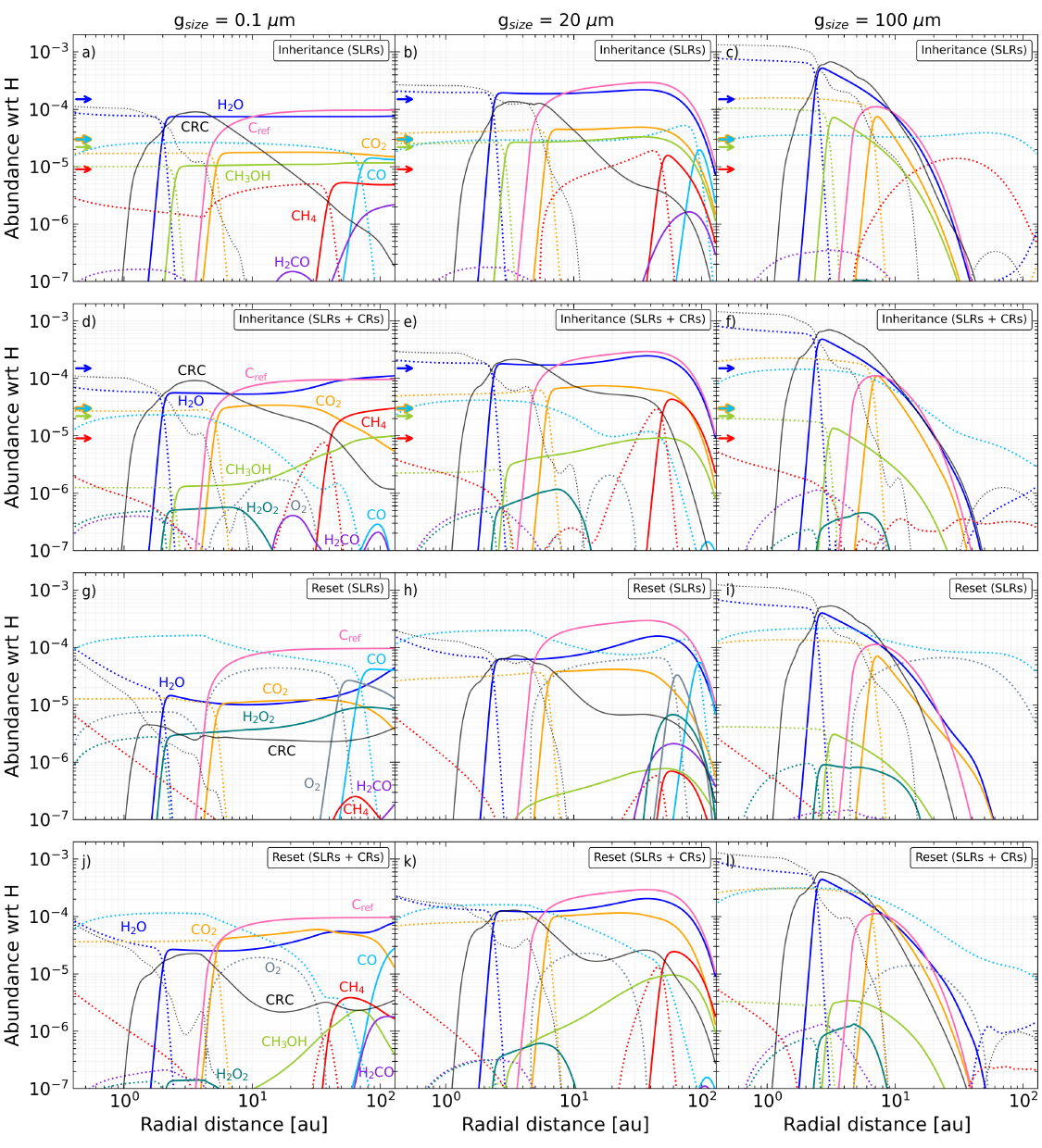}
\caption{Radial abundance profiles of the main C- and O-bearing volatile molecules in the gas phase (dotted lines) and in the ice phase (solid lines) in the disc midplane after 3~Myr of evolution with the time-dependent model (full model). The results are shown for the four chemical scenarios and the three investigated grain sizes assuming $\alpha=10^{-3}$. The horizontal small arrows indicate the initial abundances of H$_2$O, CO, CO$_2$, CH$_3$OH, and CH$_4$ in the gas phase in the inheritance scenario (see also Tab. \ref{tab:init_abund}). Note that CO and CO$_2$ have the same initial abundance. The black profile (CRC), represents the cumulative abundance of additional carbon-rich compounds, including carbon chains, hydrocarbons, nitriles, and other complex organics.} 
\label{fig:molec_main}
\end{figure*}

\begin{figure*}
\centering
\includegraphics[width = 0.9\textwidth]{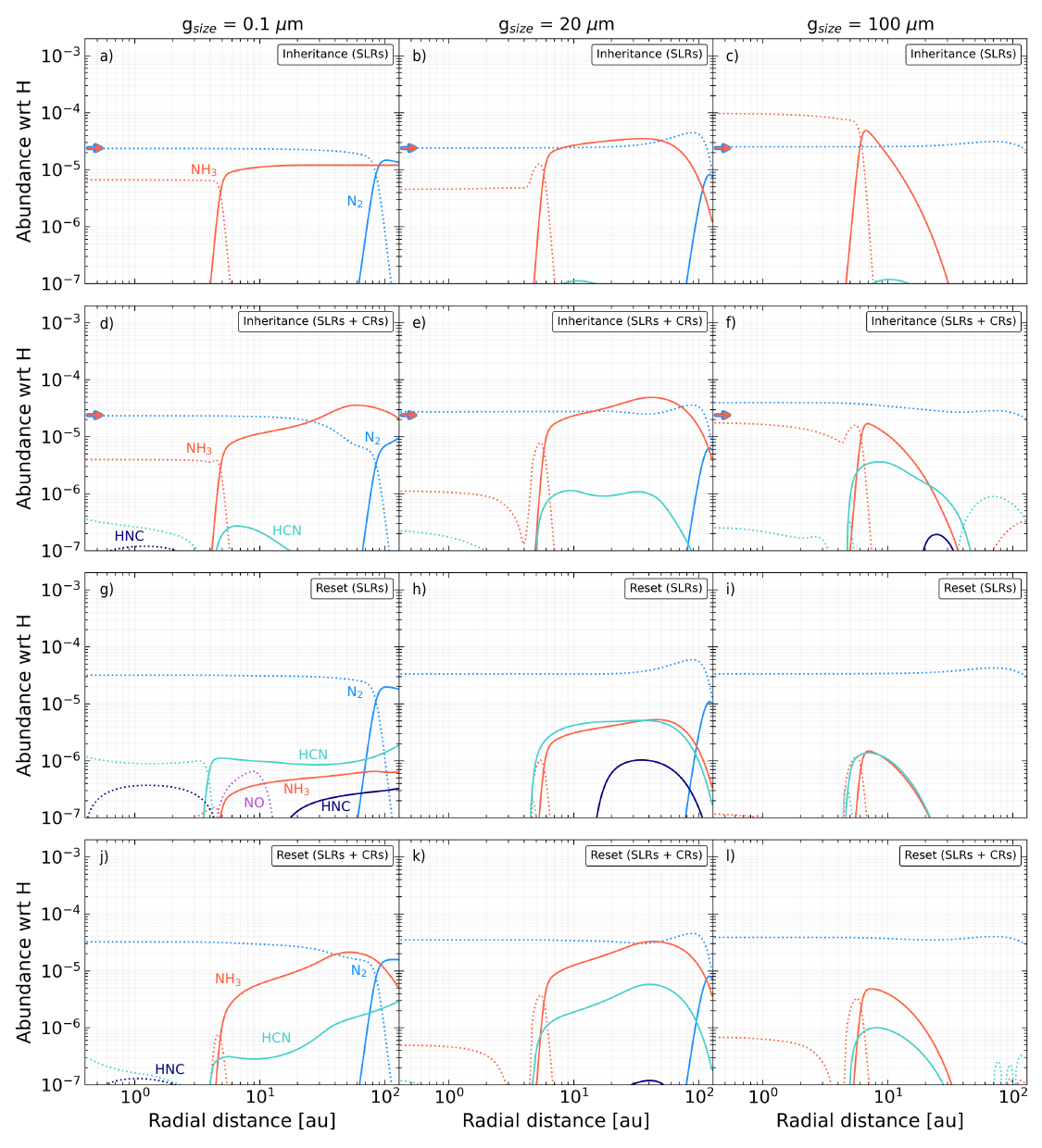}
\caption{Radial abundance profiles of the main N-bearing volatile molecules in the gas phase (dotted lines) and in the ice phase (solid lines) in the disc midplane after 3~Myr of evolution with the time-dependent model (full model). The results are shown for the four chemical scenarios and the three investigated grain sizes assuming $\alpha=10^{-3}$. The horizontal small arrows indicate the initial abundance of N$_2$ and NH$_3$ in the gas phase in the inheritance scenario (see also Tab. \ref{tab:init_abund}).} 
\label{fig:molec_nbear}
\end{figure*}

Mass transport redistributes volatiles throughout the disc, locally altering density, temperature, and ionisation rates. Its impact on disc composition depends on the competing timescales of chemical reactions and transport processes (diffusion, mixing, and radial drift), with the strongest deviations from the stationary picture expected in the inner disc, within the snowlines, and in regions of steep compositional gradients \citep{Booth2019}.

Figures \ref{fig:molec_main} and \ref{fig:molec_nbear} present the molecular composition of the disc at 3~Myr, as obtained from our fully integrated model across the four chemical scenarios and three grain sizes. Overall, our results reveal a fundamental interplay between chemistry and dynamics, demonstrating that these processes cannot be treated independently and must be modelled concurrently. This remains true even in regimes of strong gas-dust coupling and negligible radial drift, especially when a fraction of the dust has already been incorporated into larger bodies, such as planetesimals, thereby altering the elemental balance of the disc (see Sect.~\ref{sec:res_volatile_strong} for further discussion).

The most immediate chemical consequence of radial drift is the enrichment of the gas phase with volatiles inside the snowlines. The magnitude of this enrichment is governed by the relative timescales of viscous transport (Eq.\ref{eq:tvisc}) and radial drift (Eq.\ref{eq:tdrift}). When $t_{\rm drift} \ll t_{\rm visc}$, icy grains are efficiently transported inwards and cross the snowlines, where their volatile content sublimates into the gas phase. This process creates local peaks in the gas-phase abundances of volatile species near the snowlines, which are then redistributed on a viscous timescale \citep{Booth2019}. Since $t_{\rm visc}$ is longer in the outer disc, these enrichment features can still be seen at 3~Myr - for example, in the gas-phase abundances of CO, CH$_4$, and N$_2$ in the $20$\si{\micro\meter}-sized grains scenario (central column in Figs.~\ref{fig:molec_main} and \ref{fig:molec_nbear}).

Our findings indicate that even a population of relatively small grains ($\lesssim100$~\si{\micro\meter}) can drive significant volatile enrichment in the gas phase through radial drift on timescales comparable to those of planet formation (see Sect.~\ref{sec:res_volatile_weak} for details). This suggests that giant planets can accrete metal-rich gas from sublimating icy grains, even in scenarios where the bulk of pebbles (mm–cm-sized grains) has already been converted into larger bodies or accreted onto the star, leaving behind only a population of small grains in the disc. 

\subsection{Strong gas-dust coupling}\label{sec:res_volatile_strong}

The scenario with $0.1$~\si{\micro\meter}-sized grains (first column in Figs.~\ref{fig:molec_main} and \ref{fig:molec_nbear}) represents strong gas-dust coupling, where the chemical variations due to mass transport are primarily driven by viscous transport, with a negligible influence of radial drift.

Most volatile species exhibit systematically lower abundances compared to their initial values (dash-dotted lines) and the stationary solution (Fig.~\ref{fig:static_disc}). This decrease is inversely related to the volatility of the species. For example, in the inheritance-low scenario (panel \emph{a} in Figs.~\ref{fig:molec_main} and \ref{fig:molec_nbear}), the depletion at 3~Myr amounts to a factor of 2 for H$_2$O but is negligible for N$_2$. The effect is a direct consequence of inward mass transport following the early conversion of $50\%$ of the dust into planetesimals, which alters the elemental balance across the disc. As discussed in Sect.~\ref{sec:res_static_model}, planetesimal formation depletes the outer-disc ice phase of volatiles, trapping part of them in chemically inert large bodies. As a result, the remaining gas-ice mixture beyond the snowlines contains fewer volatiles than the gas originally present within the snowlines. As this gas moves inwards and is replenished by gas and ice (which sublimates at the snowlines) from the outer disc, the altered elemental balance is also transported inwards. The net effect is a gradual decrease in the gas-phase abundances of volatile species and in the gas metallicity within the snowlines (see also App.~\ref{app:A}). In the case of N$_2$, the depletion is less pronounced because it constitutes a smaller fraction of the total ice reservoir. Consequently, a smaller amount of N-bearing volatiles is sequestered in planetesimals compared to less volatile species like H$_2$O.

In the inheritance-low scenario, another notable difference from the stationary solution is that the gas-phase CH$_4$ is less depleted between 2 and 5~au due to its partial replenishment by CH$_4$ from the outer disc.

Radial diffusion and mixing tend to attenuate local inhomogeneities and smooth out abundance peaks in both the gas and ice phases. In the inheritance-high scenario (panel \emph{d}), this process counteracts the CR-driven depletion of NH$_3$ between 3 and 6~au. A similar effect is seen with H$_2$O ice, which is less depleted between 5 and 30~au. In this region, CO gas shows a gradual depletion rather than the plateau seen in the stationary solution, and O$_2$ decreases by nearly an order of magnitude. 

In the reset scenarios (panels \emph{g} and \emph{j}), CO$_2$ is strongly depleted within its snowline in the stationary solution but reaches abundances of about $10^{-5}$ (reset-low) and $4\times10^{-5}$ (reset-high) when mass transport is taken into account. In the ice phase, the most affected species is H$_2$O. In the full-model solution, its abundance remains uniform from 2~au onwards and systematically above $10^{-5}$. In contrast, in the stationary solution, it drops below $10^{-7}$ within 5~au and decreases significantly between 5 and 20~au (Sect.~\ref{sec:res_static_model}). 

\subsection{Weak gas-dust coupling}\label{sec:res_volatile_weak}

The centre and right columns of Figs.~\ref{fig:molec_main} and \ref{fig:molec_nbear} show the results for grains of size 20~\si{\micro\meter} and 100~\si{\micro\meter}, respectively, representing scenarios with non negligible radial drift. 

The abundance profiles of the ice-phase species (solid lines) closely follow the evolution of the dust surface density (Fig.~\ref{fig:surf_dens}). With 100~\si{\micro\meter}-sized grains, radial drift shrinks the dust disc to 50~au over a timescale of 2~Myr (Sect.~\ref{sec:res_transport}), leading to the complete depletion of all ice species that condense beyond the CH$_4$ snowline. 

In the gas phase (dotted lines), volatile molecules released by drifting grains in the outer disc tend to remain concentrated just inside their respective snowlines. This effect is particularly pronounced for 20~\si{\micro\meter}-sized grains at the N$_2$ snowline across all chemical scenarios (central column of Fig.~\ref{fig:molec_nbear}), as well as at the CH$_4$ and CO snowlines in the inheritance-low scenario and at the O$_2$ and CO snowlines in the reset-low scenario (panels \emph{b} and \emph{h} of Fig.~\ref{fig:molec_main}). This behaviour results from the mass flux in the disc, which approaches the inversion point near 100~au, shifting from positive to negative values and consequently slowing gas transport (Fig.~\ref{fig:acc_rate}).

Closer to the star, where the mass flux is higher, the sublimated molecules diffuse inwards on shorter timescales. This leads to a noticeable and uniform increase in the gas-phase abundances in the inner disc (within the H$_2$O snowline), compared to the scenario with 0.1~\si{\micro\meter}-sized grains. The enrichment is more pronounced for volatile molecules that condense between 2 and 10~au, compared to those condensing further out. This is because a larger fraction of their mass exists in ice form and reaches the inner disc sooner than molecules condensing at greater distances. Over a 3~Myr timescale, we find that the abundances of H$_2$O, CH$_3$OH, and NH$_3$ are the most impacted, increasing nearly twofold in the 20~\si{\micro\meter} grain-size scenario and by almost an order of magnitude in the 100~\si{\micro\meter} grain-size scenario, relative to the 0.1~\si{\micro\meter} case. In contrast, high-volatility species like CO and N$_2$ experience only a modest enrichment in the inner disc compared to their initial abundances. 

\section{Elemental ratios in an evolving disc}\label{sec:res_ratios}

Figures \ref{fig:c_o_ratio}, \ref{fig:c_n_ratio}, and \ref{fig:n_o_ratio} compare the evolution of the C/O, C/N, and N/O ratios extracted from the molecular abundance profiles returned by {\sc JADE} in the inheritance and reset scenarios with low ionisation (results for the high ionisation case are presented in Appendix~\ref{app:B}). The elemental abundances account for all molecular carriers of C, O, and N with abundances exceeding $10^{-7}$ relative to the total H nuclei. The elemental ratios in the solid phase include contributions from rocks, ice mantles, and refractory organic carbon. To generalise our results to Sun-like host stars with varying compositions, we normalise the elemental ratios to their respective stellar value and examine their radial and temporal variations as deviations from the stellar reference \citep[e.g.][]{Turrini2021, Pacetti2022}. Normalised ratios are denoted with the superscript $^*$. 

In the scenario with 0.1~\si{\micro\meter}-sized grains, the long radial drift timescale relative to the viscous timescale minimises the impact of ice sublimation on the elemental ratios. For 20~\si{\micro\meter}-sized grains, radial drift operates on a timescale of about 10~Myr, yet the resulting enrichment in the gas phase can already be noticed from 2~Myr for some elemental ratios. In the 100~\si{\micro\meter} scenario, radial drift depletes the outer disc within approximately 2~Myr, after which its impact on the elemental ratios reaches a maximum. 

\subsection{C/O ratio}\label{sec:COratio}

\begin{figure*}
\centering
\begin{subfigure}[b]{\textwidth}
    \centering
    \includegraphics[width=0.95\textwidth]{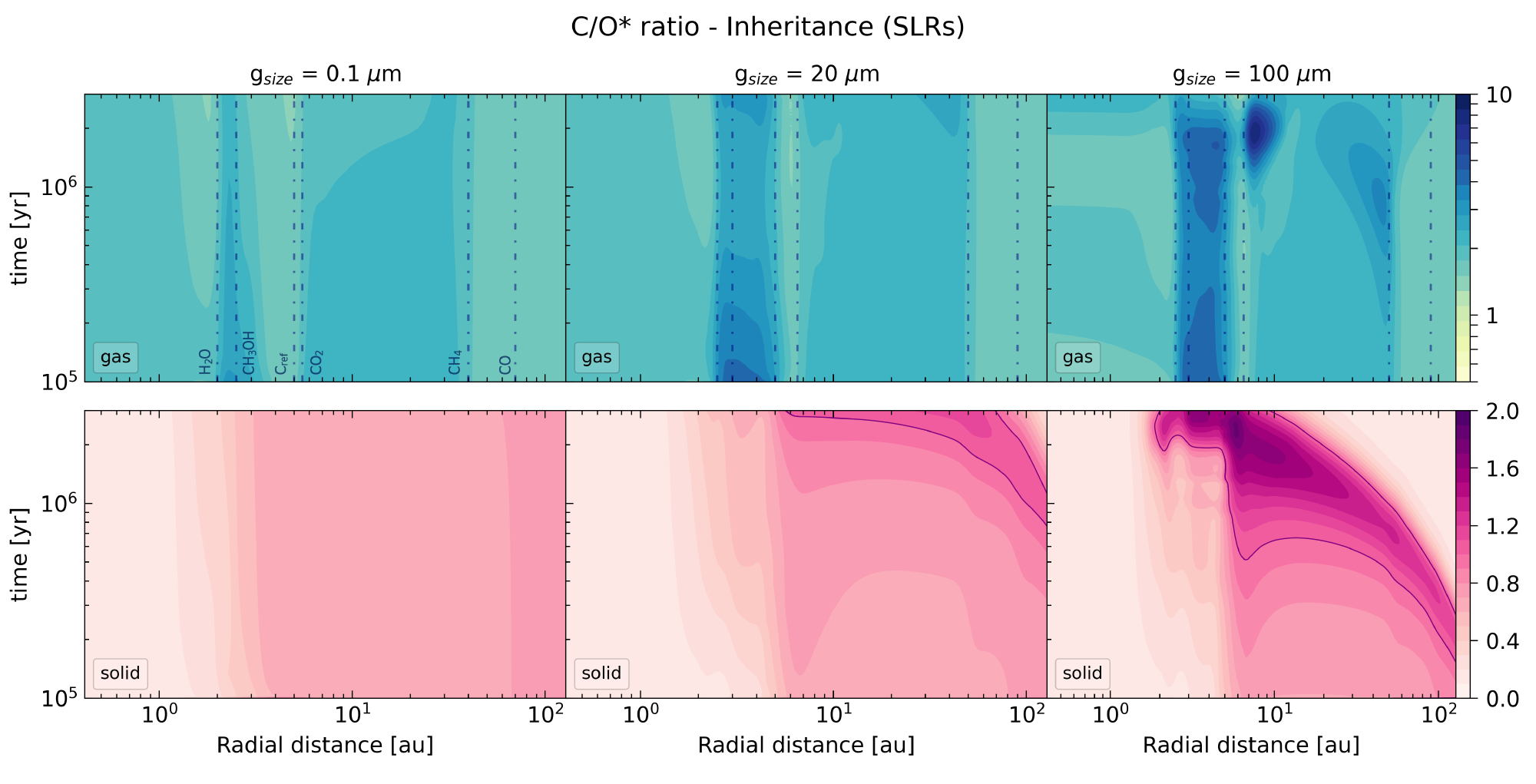}
    \caption{}
    \label{fig:c_o_ratio_inhlow}
\end{subfigure}
\vspace{2mm} 
\begin{subfigure}[b]{\textwidth}
    \centering
    \includegraphics[width=0.95\textwidth]{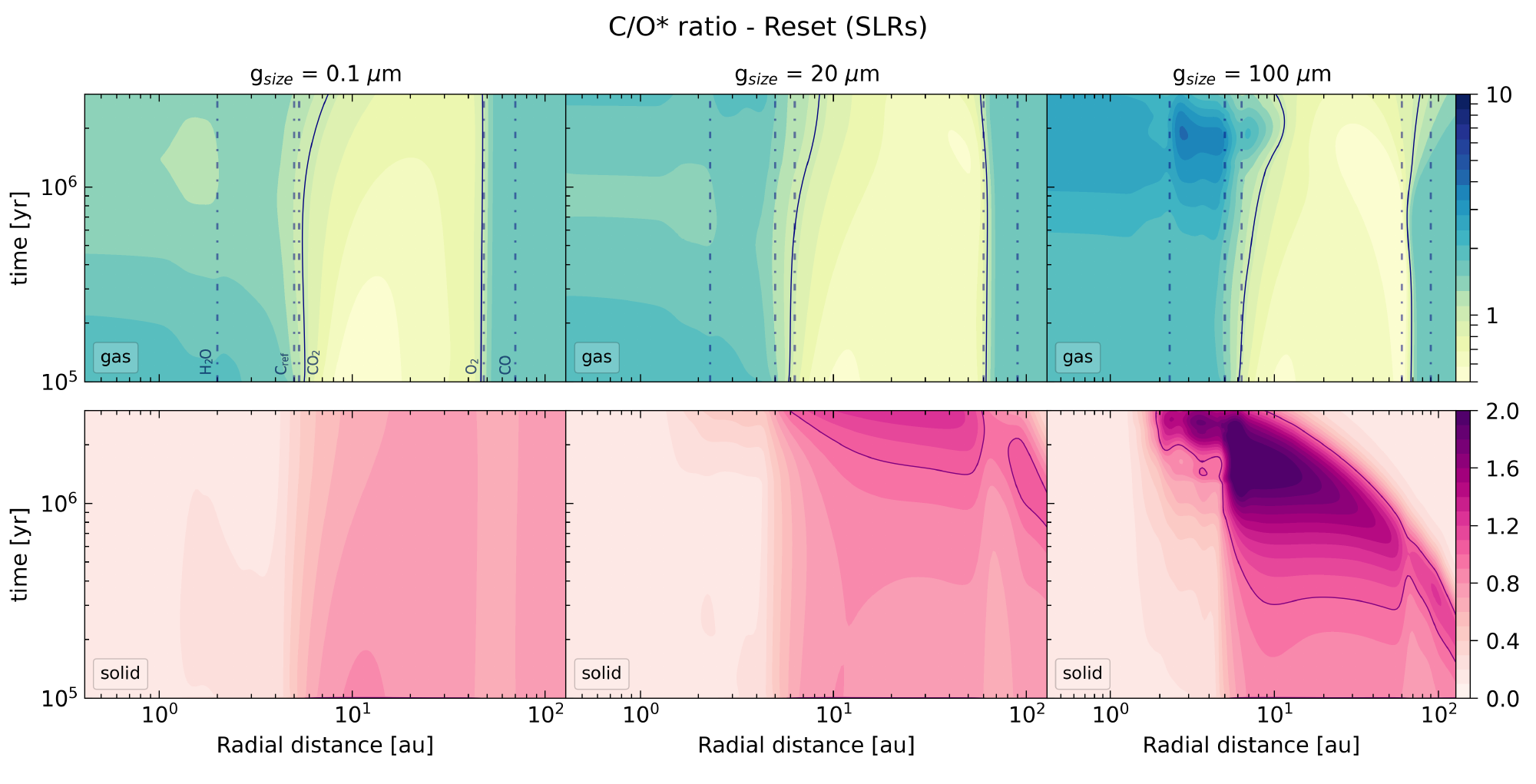}
    \caption{}
    \label{fig:c_o_ratio_reslow}
\end{subfigure}
\caption{C/O ratio in the disc as a function of distance and time for the inheritance (panel \emph{a}) and reset (panel \emph{b}) scenarios with low ionisation. In each panel, the top row represents the gas phase, while the bottom row shows the solid phase (including rocks, ice mantles, and refractory organic carbon). All values are normalised to the stellar C/O ratio ($\sim 0.59$), with normalisation indicated by the superscript $^*$. The dark contour represents a value of~1, corresponding to the stellar ratio. Vertical lines indicate the approximate locations of key snowlines at 2~Myr. Molecules are labelled only in the top-left plot of each panel but follow the same order in the other plots.}
\label{fig:c_o_ratio}
\end{figure*}

The combined effects of chemical processes and mass transport in the disc produce local deviations in the C/O ratio (Fig.~\ref{fig:c_o_ratio}) from the stellar value. These deviations increase with grain size and become significant even for sub-millimetre-sized grains. The inheritance-low case with 0.1~\si{\micro\meter}-sized grains (top-left plot in panel \emph{a}) is the least affected by disc evolution, with deviations from the stellar value remaining within $20\%$, except between 1.5 and 3~au, where they reach peaks of $50\%$. 

In this scenario, our model does not reproduce the classical monotonic trend in the gas-phase C/O ratio proposed by \citet{Oberg2011}. The inclusion of rocks, refractory organic carbon, and a broader range of C- and O-bearing volatile molecules results in a more complex radial structure for the C/O ratio \citep{Turrini2023}, which is then further shaped by disc evolution.  

The two different sets of initial chemical conditions leave distinct signatures. In the inheritance scenario (panel \emph{a}), the gas-phase C/O* ratio is always superstellar, reaching peak values up to five times the stellar ratio in discs with 20~\si{\micro\meter}-sized grains and up to eight times the stellar ratio in discs with 100~\si{\micro\meter}-sized grains. Assuming a solar composition and thus a C/O ratio of 0.59 \citep{Asplund2021}, these values correspond to C/O ratios of up to 2.95 and 4.72, respectively. In contrast, the gas-phase C/O* ratio in the reset scenario (panel \emph{b}) is substellar across a wide region of the disc, extending from the CO$_2$ to the O$_2$ snowlines ($\sim 6-50$~au). This substellar trend is due to O$_2$ entering the gas phase and enriching it with oxygen relative to carbon (see Sect.~\ref{sec:res_static_model}). This behaviour of the C/O* ratio highlights a key difference between planet-forming environments that have inherited their composition from the prestellar phase and those that have undergone a chemical reset, extending the result found by \citet{Eistrup2016} for static discs.

Temporal changes in the solid-phase C/O* ratio (bottom row in panels \emph{a} and \emph{b}) are mainly driven by the radial drift of dust grains. These changes are minimal for submicrometre grains but become substantial for larger grains. The accumulation of carbon-rich grains near the CO$_2$ and C$_{ref}$ snowlines increases the solid-phase C/O* ratio to twice the stellar value after 1~Myr in the disc with 100~\si{\micro\meter}-sized grains. On the solar scale, this would correspond to a C/O ratio of approximately 1.2, reached within a relatively narrow region where the effects of radial drift are most pronounced. 

In the gas phase (top row in panels \emph{a} and \emph{b}), the C/O* ratio exhibits more substantial variations over time. In most grain-size scenarios, we can distinguish two regimes: before and after radial drift becomes efficient. In the first regime ($t\lesssim 0.5$~Myr), the evolution of the C/O* ratio is concentrated in the inner disc, within the C$_{ref}$ snowline. As discussed in Sect.~\ref{sec:res_volatile_strong}, one of the mechanisms driving chemical evolution in the inner disc is the inward diffusion of gas depleted in volatiles after the formation of planetesimals (Sect.~\ref{sec:res_volatile_strong}). The impact of this process on the gas-phase C/O* ratio depends on the relative amounts of C and O sequestered into planetesimals beyond the snowlines. In the inheritance scenario (top row in panel \emph{a}), the C and O depletions nearly offset each other within the H$_2$O snowline, limiting the temporal variations of the C/O* ratio to less than 10$\%$ (the region between the H$_2$O and C$_{ref}$ snowlines is discussed in detail later). In contrast, in the reset scenario (top row in panel \emph{b}), the high abundance of molecular oxygen in the gas phase in the disc region where most volatiles condense as ice results in a smaller fraction of oxygen being incorporated into planetesimals. Consequently, inward diffusion causes the inner-disc gas to become more depleted in carbon than oxygen, leading to a declining C/O* ratio within the C$_{ref}$ snowline. This effect is particularly pronounced in discs with 0.1 and 20~\si{\micro\meter}-sized grains, where the C/O* ratio decreases by up to $20\%$. 

From about 1~Myr, the evolution of the gas-phase C/O* ratio is primarily driven by the sublimation of drifting icy grains and becomes most evident near the snowlines. In both chemical scenarios, local peaks in the C/O* ratio are especially pronounced in the disc with 100~\si{\micro\meter}-sized grains, particularly within the C$_{ref}$ snowline, where the C/O* ratio can reach up to five times the stellar value. Ice sublimation is also responsible for the increased C/O* ratio within the CH$_4$ snowline in the inheritance scenario and the decreased C/O* ratio within the O$_2$ snowline in the reset scenario. Our results also reveal a narrow region between 7 and 10~au, just beyond the CO$_2$ snowline, where the gas-phase C/O* ratio reaches values up to eight times the stellar value. This elevated C/O* ratio is driven by an enhanced formation of carbon chains and hydrocarbons (e.g. C$_2$ and H$_2$CCC) in the gas phase between 1 and 3~Myr, peaking at 1.8~Myr. In the inheritance scenario, this process results in a total carbon abundance 4.5 times higher than that of oxygen. 

\begin{figure}
\centering
\includegraphics[width = \columnwidth]{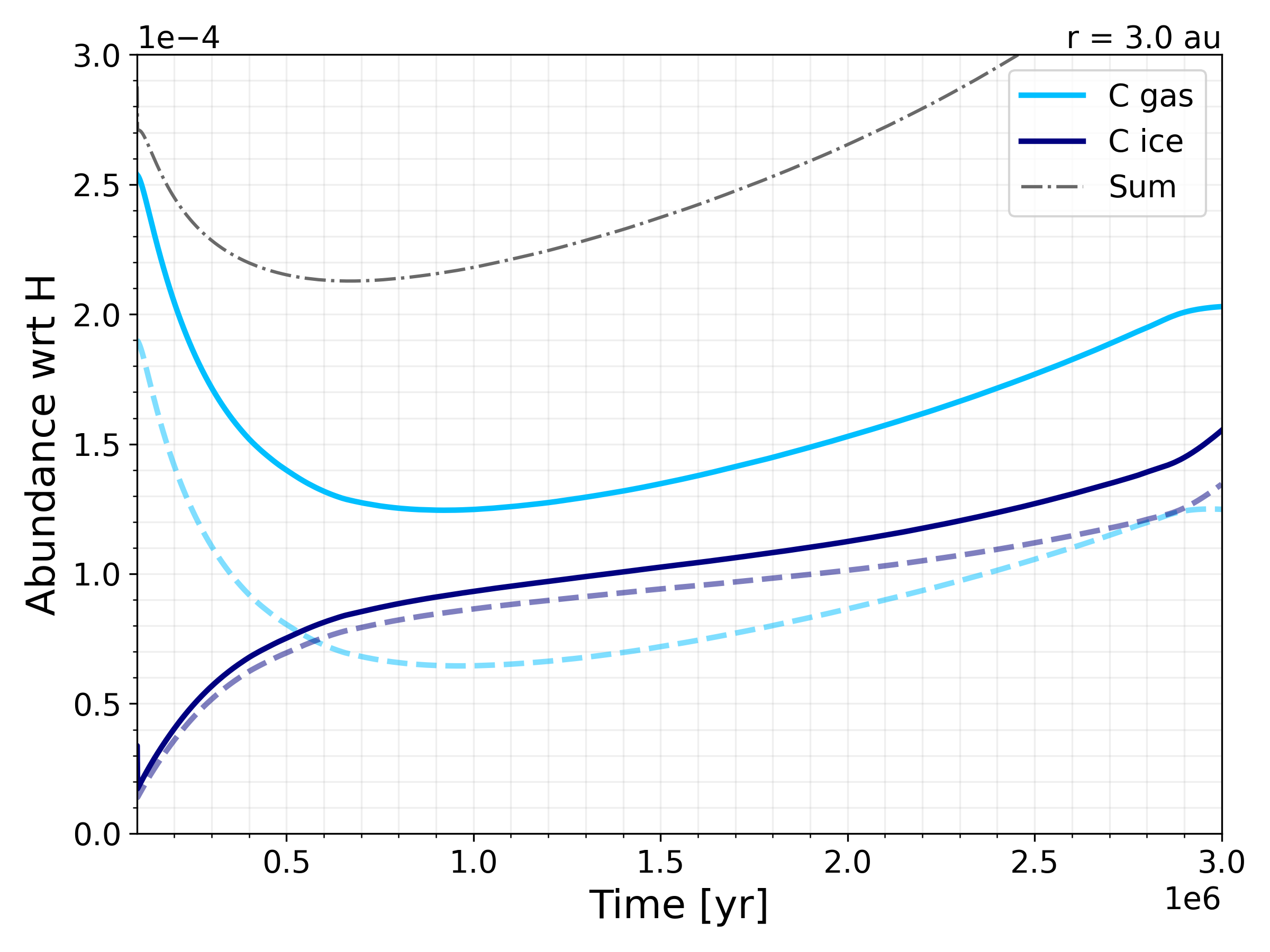}
\caption{Evolution of carbon abundance at 3~au in the inheritance-low scenario, assuming dust grains with a radius of 20~\si{\micro\meter}. The solid lines represent the total carbon in the gas (light blue) and ice (dark blue) phases, while the dashed lines isolate the contribution of CRC. The grey dash-dotted line indicates the combined total of all contributions.} 

\label{fig:Cevol_20mic}
\end{figure}

In the inheritance scenario, we identify a chemically distinct region between the H$_2$O and C$_{ref}$ snowlines. Here, the gas-phase C/O* ratio decreases in the first 0.5~Myr across all grain size scenarios, with steeper variations corresponding to larger grain sizes, and then increases again thereafter. The decreasing C/O* ratio results from the freeze-out of the chemical products of C$_{ref}$ reprocessing within its snowline. This mechanism is illustrated in Fig.~\ref{fig:Cevol_20mic} for the disc with 20~\si{\micro\meter}-sized grains. The plot shows the evolution of the carbon abundance at 3~au. The solid curves represent all carbon carriers that reach abundances greater than $10^{-7}$ during the disc's evolution, while the dashed curves highlight the contributions of CRC. In the inheritance scenarios, these compounds are efficiently produced in the gas phase from the atomic carbon released within the C$_{ref}$ snowline (see Sect.~\ref{sec:res_static_model}). During the first 0.5~Myr, a fraction of these species condenses as ice, leading to a decrease in gas-phase carbon and a corresponding increase in ice-phase carbon, which results in a lower gas-phase C/O* ratio. The total carbon abundance (grey dash-dotted curve) decreases as gas diffuses inwards and is replaced by material that is depleted in carbon after the formation of planetesimals. After 1~Myr, radial drift becomes efficient, causing the carbon abundance and the C/O* ratio to increase again. 

Similar global differences between the inheritance and reset scenarios are observed under high ionisation (see Fig.~\ref{fig:c_o_ratio_high}), although radial drift combined with more active chemistry leads to a more moderate increase in the gas-phase C/O ratio within 10~au, with values reaching up to three times the stellar C/O ratio in the disc with 100~\si{\micro\meter}-sized grains.

\subsection{C/N ratio}\label{sec:CN_ratio}

\begin{figure*}
\centering
\begin{subfigure}[b]{\textwidth}
    \centering
    \includegraphics[width=0.95\textwidth]{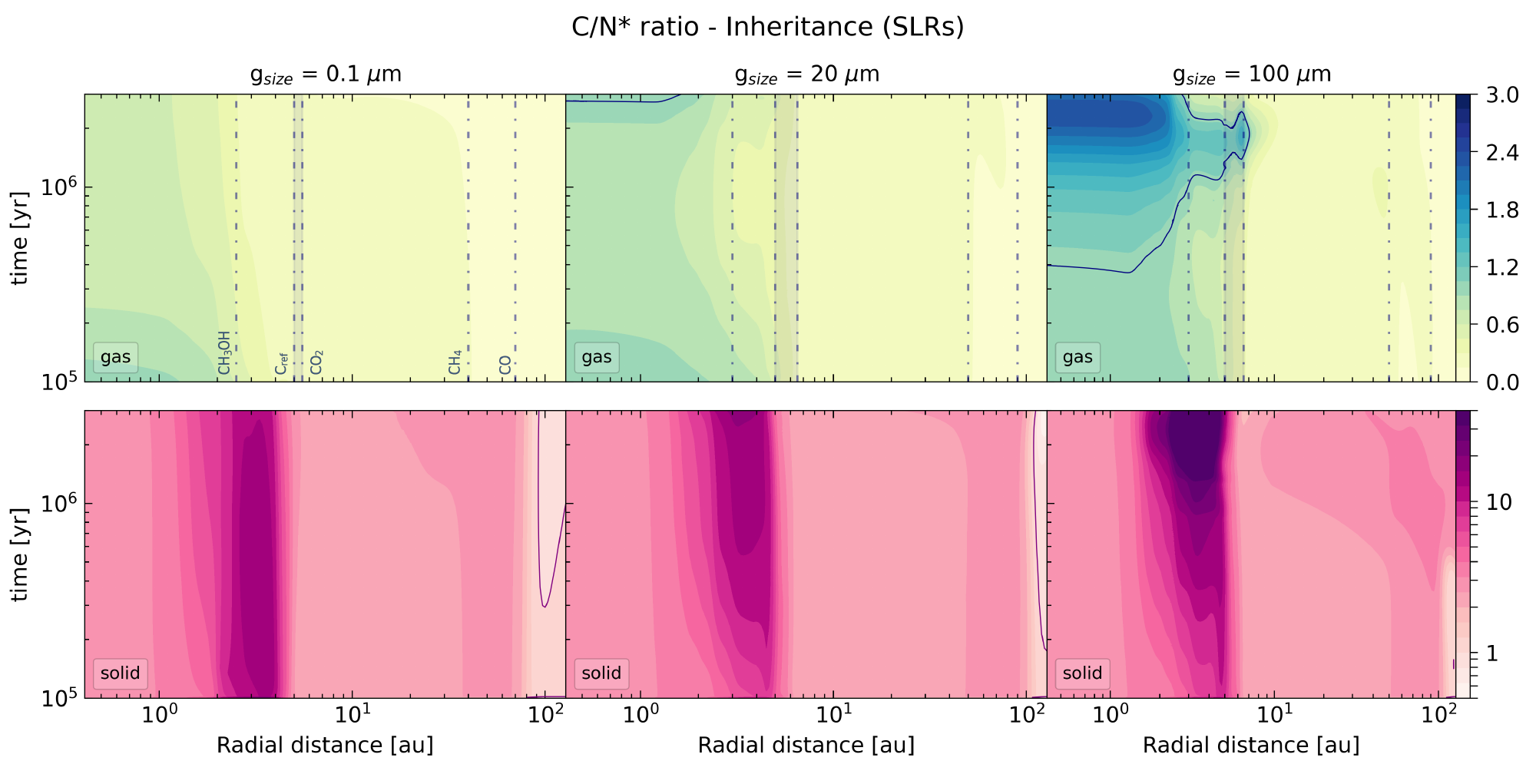}
    \caption{}
    \label{fig:c_n_ratio_inhlow}
\end{subfigure}
\vspace{2mm} 
\begin{subfigure}[b]{\textwidth}
    \centering
    \includegraphics[width=0.95\textwidth]{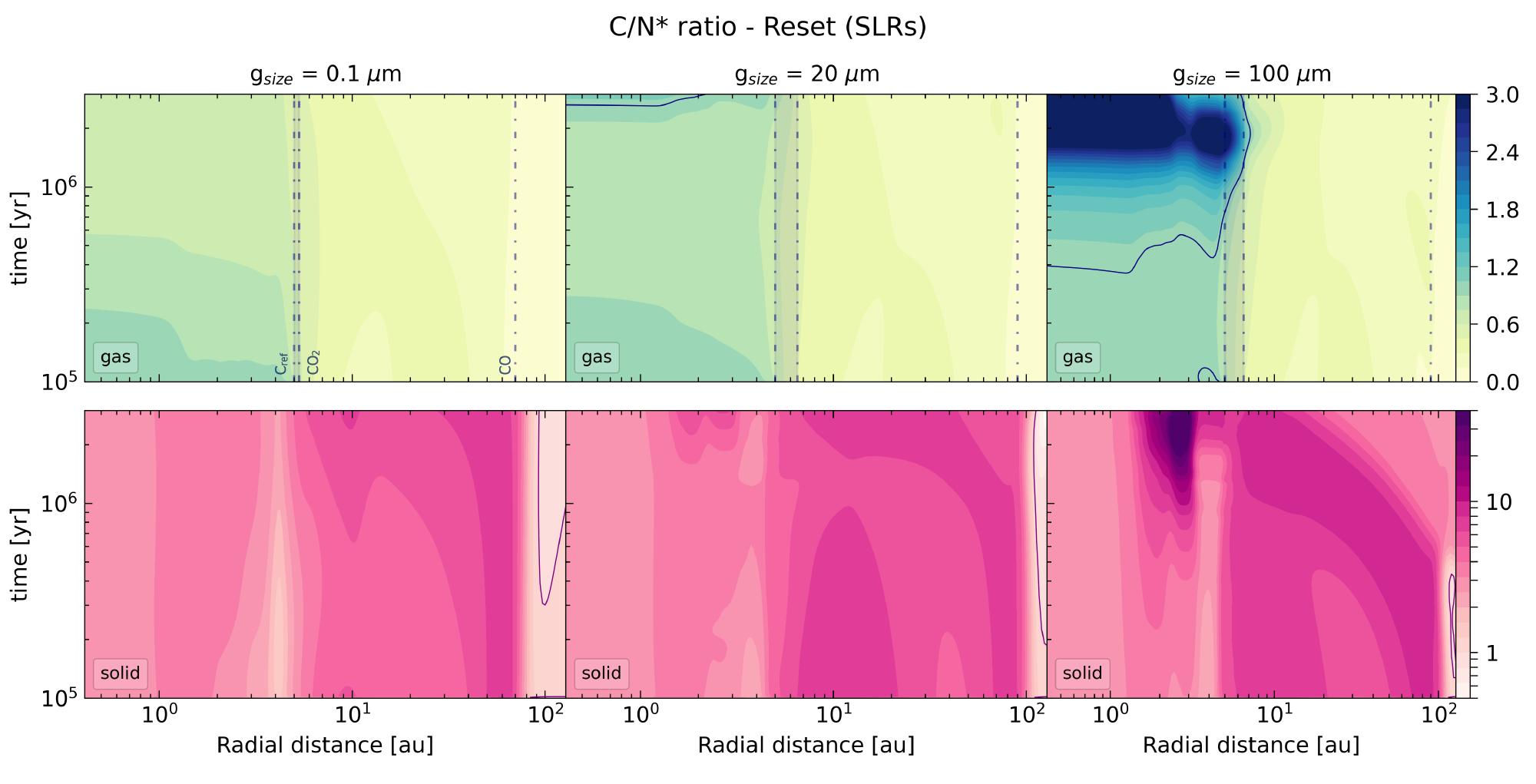}
    \caption{}
    \label{fig:c_n_ratio_reslow}
\end{subfigure}
\caption{C/N ratio in the disc as a function of distance and time for the inheritance (panel \emph{a}) and reset (panel \emph{b}) scenarios with low ionisation. In each panel, the top row represents the gas phase, while the bottom row shows the solid phase (including rocks, ice mantles, and refractory organic carbon). All values are normalised to the stellar C/N ratio ($\sim 4.27$), with normalisation indicated by the superscript $^*$. The dark contour represents a value of~1, corresponding to the stellar ratio. Vertical lines indicate the approximate locations of key snowlines at 2~Myr. Molecules are labelled only in the top-left plot of each panel but follow the same order in the other plots. For clarity, the NH$_3$ snowline in both scenarios is not shown but lies within the shaded region between the C$_{ref}$ and CO$_2$ snowlines.}
\label{fig:c_n_ratio}
\end{figure*}

The C/N* ratio (Fig.~\ref{fig:c_n_ratio}) appears to be a weaker diagnostic of the disc’s initial chemical conditions (inheritance vs. reset) and ionisation level (high vs. low) than the C/O* ratio. However, it exhibits steeper and more radially extended temporal variations in the inner disc, inward of the CO$_2$ snowline, particularly in the 100~\si{\micro\meter} grain size scenario, making it a more effective tracer of ice sublimation driven by radial drift.

The distinct behaviour of the C/N* ratio originates from the higher volatility of nitrogen compared to carbon and oxygen. In both the inheritance and reset scenarios, approximately $60\%$ of the initial volatile nitrogen reacts to form N$_2$ early in the disc's evolution, with its abundance remaining relatively stable throughout most of the disc. Variations in N$_2$ abundance are primarily driven by the sublimation of ices drifting from beyond the N$_2$ snowline, though this effect is largely confined to the outer disc, where the N$_2$ snowline lies (see Fig.~\ref{fig:molec_nbear}). The significant retention of nitrogen in the gas phase results in the gas-phase C/N* ratio (top row in panels \emph{a} and \emph{b}) exhibiting less pronounced variations than the C/O* ratio, ranging from substellar values to three times the stellar value over the course of disc evolution. In contrast, the C/N* ratio in the solid phase (bottom row in panels \emph{a} and \emph{b}) exhibits a much wider dynamic range across the snowlines, due to the limited nitrogen content in the ice phase. 

In the gas phase, the C/N* ratio shows limited sensitivity to the initial chemical conditions. In both the inheritance and reset scenarios, two main compositional regions can be identified, separated by the CO$_2$ snowline: beyond the snowline, where the C/N* ratio is substellar, and within the snowline, where it becomes predominantly superstellar.

Focusing on the inner region, the same considerations for the C/O* ratio apply: the early-stage evolution of the gas-phase C/N* ratio is dominated by the inward diffusion of volatile-depleted gas, while radial drift becomes important after approximately 1~Myr. However, unlike the C/O* ratio, the C/N* ratio traces the effects of ice sublimation even within the H$_2$O snowline. In the case of strong radial drift (e.g. 100~\si{\micro\meter} grain size scenario), the C/N* ratio in this region reaches peak values up to three times the stellar value (see below for more details). In contrast, for the C/O* ratio, the enrichment is mostly confined between the H$_2$O and the C$_{ref}$ snowlines. This different behaviour arises from the higher volatility of nitrogen relative to carbon, which makes the C/N* ratio more sensitive to variations in carbon abundance compared to the C/O* ratio. 

For large grains in the inheritance scenario (top-right plot in panel \emph{a}), the late-stage enrichment of the C/N* ratio is attenuated by the release of NH$_3$ at its snowline. Ice sublimation within 6~au causes the gas-phase NH$_3$ abundance to exceed that of N$_2$, peaking at $1.4 \times 10^{-4}$ around 2~Myr and gradually declining thereafter (top-right plot in Fig.~\ref{fig:molec_nbear}). This release of NH$_3$ reduces the gas-phase C/N* ratio, producing a distinct dip between the C$_{ref}$ and CO$_2$ snowlines at approximately 2~Myr. As carbon continues to sublimate within the C$_{ref}$ and CH$_3$OH snowlines, the C/N* ratio rises again, eventually reaching values up to twice the stellar value. In the reset scenario (top-right plot in panel \emph{b}), on the other hand, N$_2$ remains about an order of magnitude more abundant than other nitrogen-bearing species, resulting in a consistently higher C/N* ratio within the CO$_2$ snowline, with up to a threefold enrichment at later stages. Compared to the inheritance scenario, the region inward of the CO$_2$ snowline appears more uniformly enriched, with the main exception occurring around 3~au. At this radius, the C/N* ratio evolves more slowly due to the efficient production of cyanoacetylene (HC$_3$N) and methylenimine (CH$_2$NH) in the gas phase. The abundance of these two molecules starts to increase after about 0.5~Myr and reaches its peak between 1.9 and 2.4~Myr, with HC$_3$N approaching N$_2$ levels and CH$_2$NH reaching $1.4 \times 10^{-5}$. Our results suggest that these molecules may act as significant nitrogen carriers in carbon-rich environments, though further investigation is required to confirm this possibility. Similar enrichments of C- and N-bearing organics in carbon-rich gas have also been reported in other recent models \citep[e.g.][]{Wei2019, Legal2019}, suggesting that this chemical behaviour may be a general feature of carbon-rich disc environments.

The global trends discussed above are preserved under high ionisation (see Fig.~\ref{fig:c_n_ratio_high}), although we find systematically lower gas-phase C/N ratios beyond 30~au in both inheritance and reset scenarios, driven by enhanced CO depletion (see also Sect.~\ref{sec:res_vermetti}).

\subsection{N/O ratio}\label{sec:NO_ratio}

\begin{figure*}
\centering
\begin{subfigure}[b]{\textwidth}
    \centering
    \includegraphics[width=0.95\textwidth]{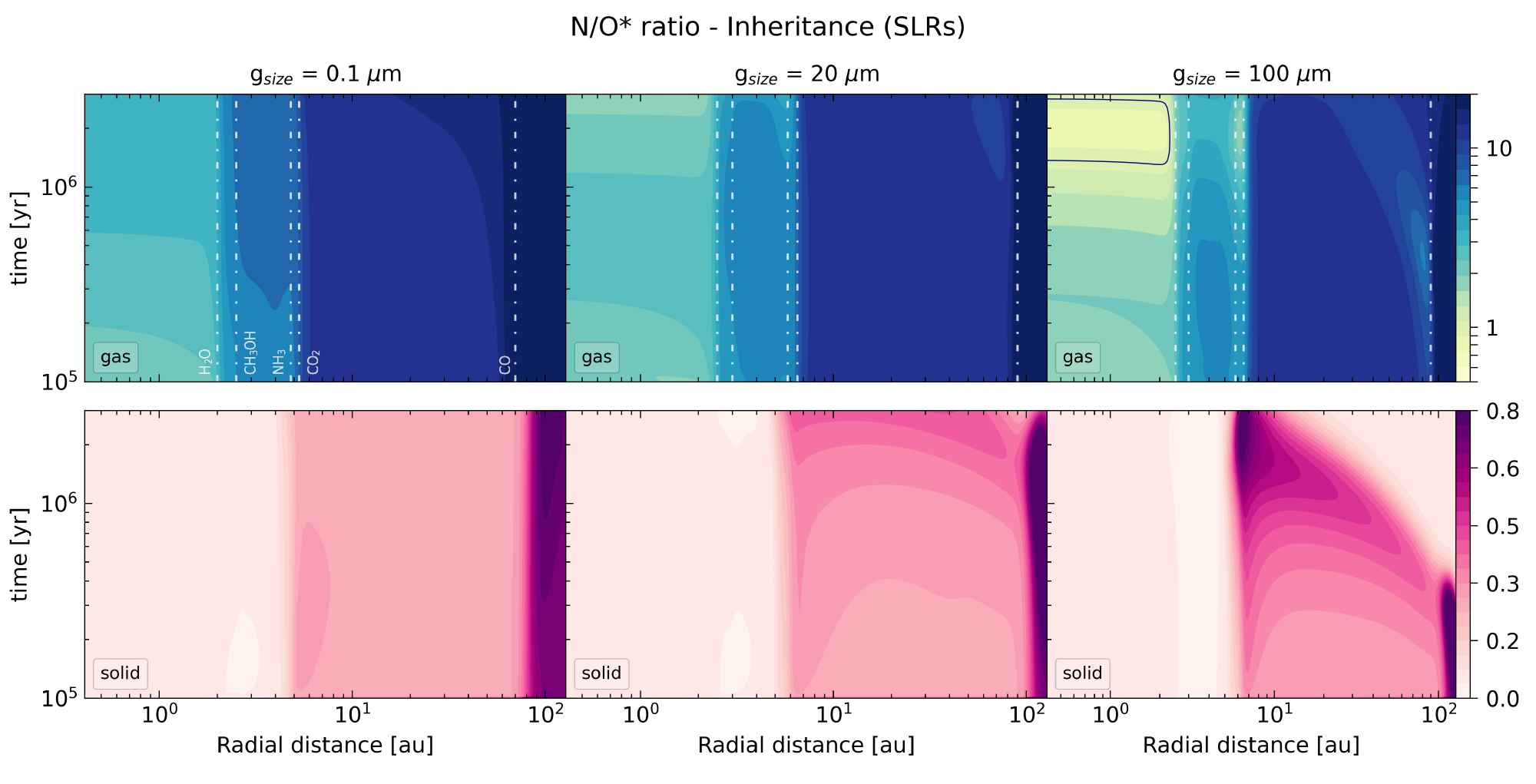}
    \caption{}
    \label{fig:n_o_ratio_inhlow}
\end{subfigure}
\vspace{2mm} 
\begin{subfigure}[b]{\textwidth}
    \centering
    \includegraphics[width=0.95\textwidth]{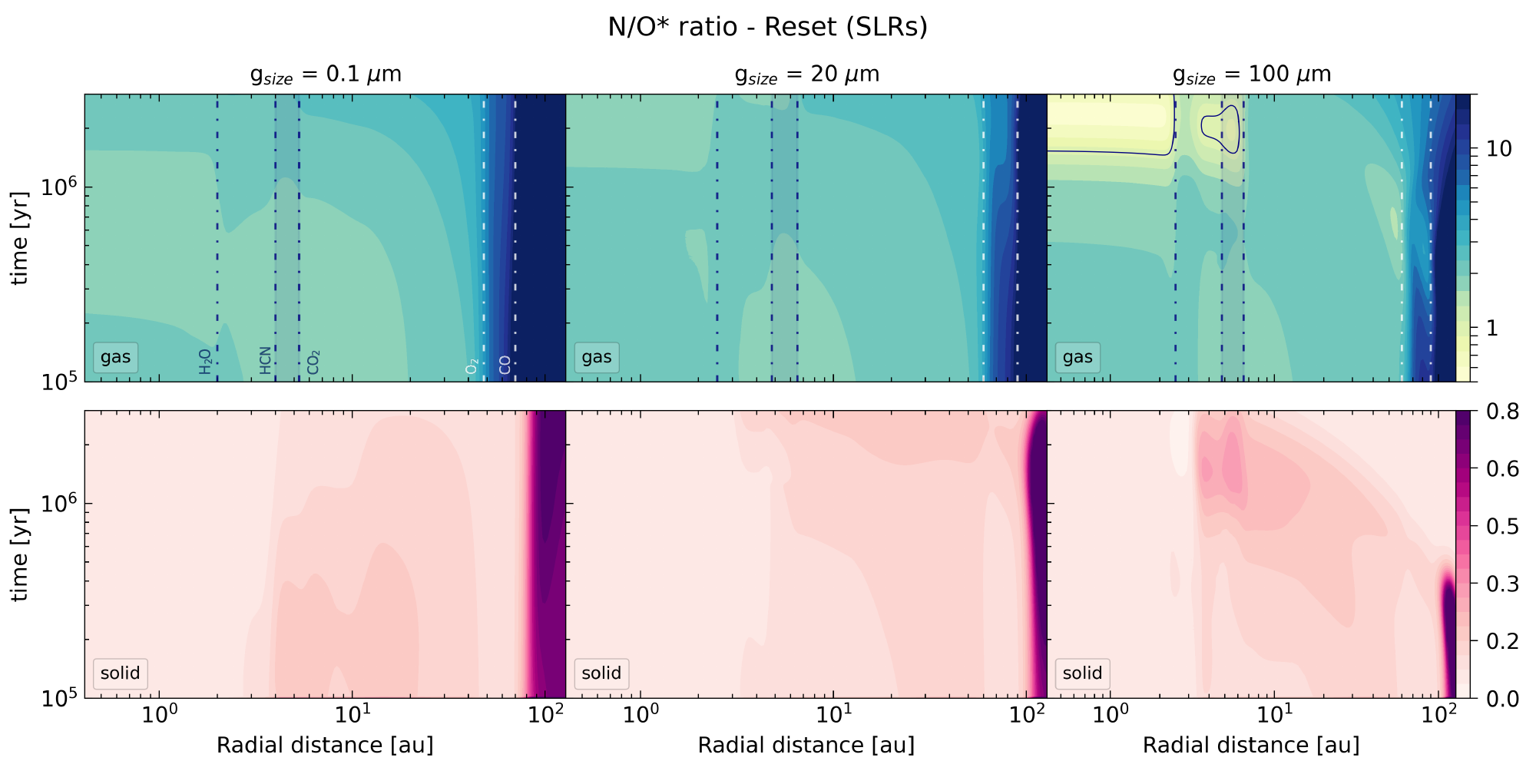}
    \caption{}
    \label{fig:n_o_ratio_reslow}
\end{subfigure}
\caption{N/O ratio in the disc as a function of distance and time for the inheritance (panel \emph{a}) and reset (panel \emph{b}) scenarios with low ionisation. In each panel, the top row represents the gas phase, while the bottom row shows the solid phase (including rocks, ice mantles, and refractory organic carbon). All values are normalised to the stellar N/O ratio ($\sim 0.14$), with normalisation indicated by the superscript $^*$. The dark contour represents a value of~1, corresponding to the stellar ratio. Vertical lines indicate the approximate locations of key snowlines at 2~Myr. Molecules are labelled only in the top-left plot of each panel but follow the same order in the other plots. For clarity, the NH$_3$ snowline in the reset scenario is not shown but lies within the shaded region between the HCN and CO$_2$ snowlines.}
\label{fig:n_o_ratio}
\end{figure*}

The N/O* ratio (Fig.~\ref{fig:n_o_ratio}) is sensitive to both the initial chemical conditions and mass transport effects in the disc, but its diagnostic power varies across different regions. Significant differences between the inheritance and reset scenarios only arise beyond the H$_2$O snowline. Conversely, the influence of mass transport is most pronounced within the CO$_2$ snowline, where the radial drift leaves a clear imprint in the gas phase after about 1~Myr for grain sizes around 100~\si{\micro\meter}.

Since nitrogen is predominantly present in the gas phase as N$_2$, while about $50\%$ of oxygen is initially locked in rocks, the solid-phase N/O* ratio (bottom row in both panels) remains strongly substellar throughout most of the disc. The N/O* ratio approaches the stellar value only beyond the N$_2$ snowline or in regions strongly enriched in nitrogen ice, such as near the NH$_3$ snowline under strong radial drift conditions (e.g. 100~\si{\micro\meter} grain size scenario) in the inheritance scenario. In contrast, the gas-phase N/O* ratio (top row in both panels) exhibits a much wider dynamic range, with values ranging from substellar to strongly superstellar values. 

In the inheritance scenario (top row in panel \emph{a}), the gas-phase N/O* ratio evolves radially in response to the condensation of O-bearing molecules and NH$_3$. Moving outwards from the star, the ratio increases from about stellar values within the H$_2$O snowline to about ten times the stellar value beyond the CO$_2$ snowline. Beyond the CO and N$_2$ snowlines, the N/O* ratio rises even further, driven by the strong depletion of O- and N-bearing species in the gas phase of the cold outer disc.

In the reset scenario (top row in panel \emph{b}), the N/O* ratio exhibits more limited radial variation, remaining below twice the stellar value throughout most of the disc. This is because oxygen is predominantly carried by CO and O$_2$, which maintain similar abundances in the gas phase and condense only in the cold outer disc. In this scenario, the depletion of NH$_3$ and other minor N-bearing species further reduces the influence of snowlines on the N/O* ratio.

Temporal variations in the gas-phase N/O* ratio are most pronounced inside the CO$_2$ snowline and in the outer disc. Under strong radial drift (e.g. 100~\si{\micro\meter} grain size scenario), the sublimation of oxygen-rich icy grains causes the gas-phase N/O* ratio to rapidly drop to substellar values after approximately 1~Myr in both chemical scenarios. In the inheritance scenario, this trend is partially offset by the sublimation of NH$_3$-rich grains, keeping the ratio superstellar between the H$_2$O and NH$_3$ snowlines. Additional variations occur in the outer disc near the O$_2$ and N$_2$ snowlines, where radial drift overtakes viscous transport on a timescale of less than 1~Myr. 

At high ionisation (see Fig.~\ref{fig:n_o_ratio_high}), similar global trends are recovered, though the gas-phase N/O ratio exhibits more pronounced temporal variations beyond 10~au, primarily due to enhanced CO depletion.

\subsection{Multi-Element Analysis of Disc Composition}\label{sec:res_vermetti}

\begin{figure*}
\centering
\includegraphics[width=0.95\textwidth]{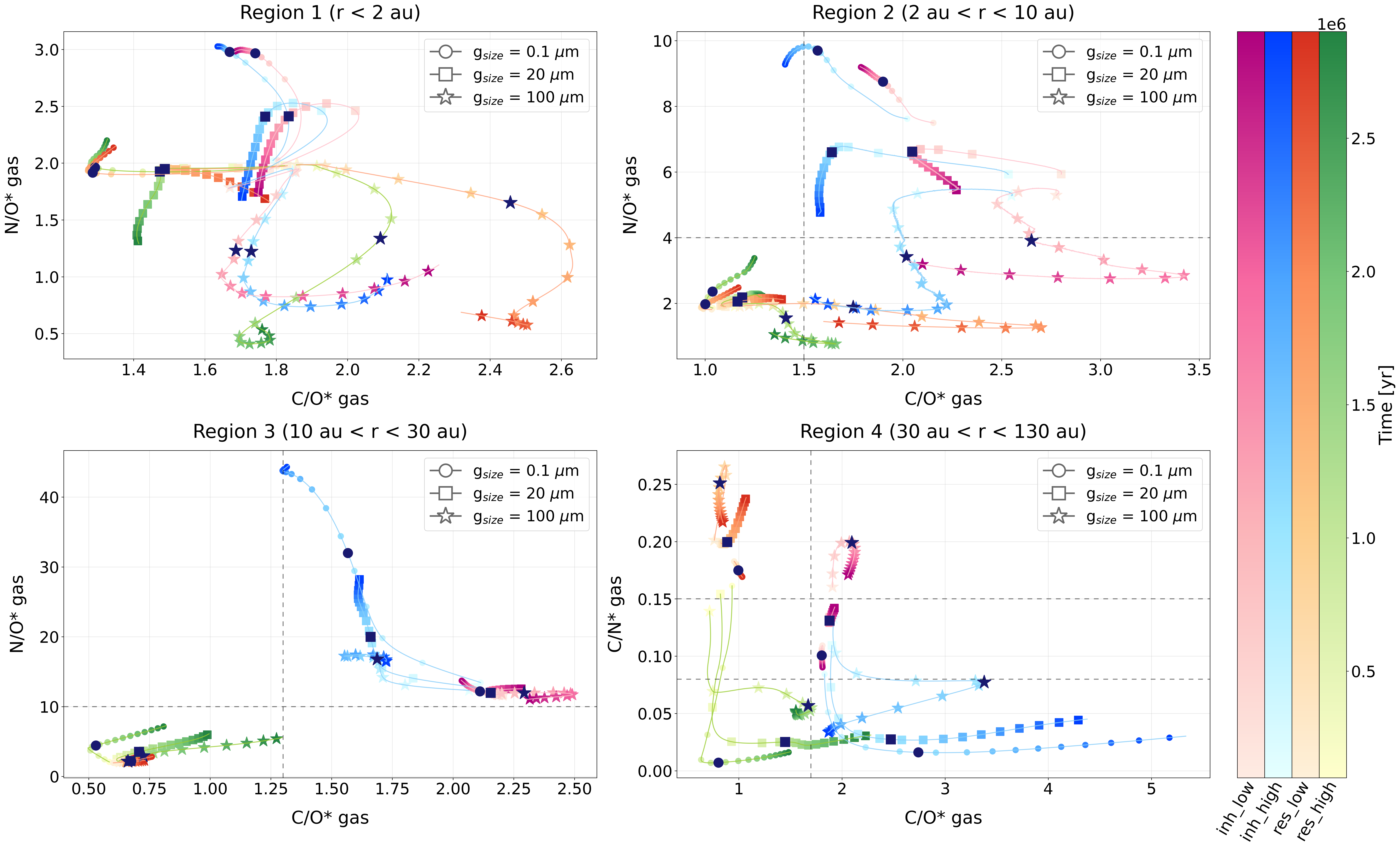}
\caption{Evolutionary tracks of the elemental ratios over 3~Myr across all considered scenarios. Ratios are shown in pairs across four key compositional regions: within the H$_2$O snowline (Region 1, top-left), between the H$_2$O and CO$_2$ snowlines (Region 2, top-right), between the CO$_2$ and CH$_4$ snowlines (Region 3, bottom-left), and beyond the CH$_4$ snowline (Region 4, bottom-right). Values represent gas-phase elemental ratios, calculated from total elemental abundances averaged over the radial extent of each region and weighted by the surface density of the gas. The colour bars indicate the four chemical scenarios, with time progressing from lighter to darker shades. The three different markers represent the selected grain sizes and are placed along each track every $200\,000$~yr. The dark marker on each track denotes 1 Myr. Dashed grey lines highlight regions of the parameter space where pairwise comparisons of the elemental ratios provide constraints on specific scenarios or subsets of scenarios.} 
\label{fig:vermetti}
\end{figure*}

In this section, we present a pairwise comparison of elemental ratios and show how a multi-element approach helps to identify signatures of disc evolution across different regions of the disc.

Figure \ref{fig:vermetti} compares pairs of normalised elemental ratios in four key compositional regions, delimited by the snowlines of H$_2$O, CO$_2$, and CH$_4$. Each panel shows the evolution of the ratios over 3~Myr for all simulated scenarios. The values represent the gas-phase elemental ratios at different time steps calculated from the total abundances of C, O, and N averaged over the radial extent of each region and weighted by the surface density profile of the gas.

Distinct chemical scenarios generally do not leave clear signatures within the H$_2$O snowline (Region 1, top-left), where all ices are sublimated, especially at early stages ($<1$~Myr). The evolutionary paths of the N/O* and C/O* ratios often overlap, ranging from substellar to superstellar N/O* ratios and from stellar to superstellar C/O* ratios. However, while superstellar C/O* and N/O* ratios remain highly degenerate, C/O* ratios near stellar values are more typical of later stages ($>1$~Myr) in reset scenarios with negligible radial drift (e.g. 0.1~\si{\micro\meter}-sized grains). In contrast, substellar N/O* ratios emerge only in scenarios with efficient radial drift (e.g. 100~\si{\micro\meter}-sized grains after 1~Myr), where a larger fraction of oxygen is released into the gas phase compared to nitrogen. 

The dynamic range of N/O* and C/O* variations broadens between the H$_2$O and CO$_2$ snowlines (Region 2, top-right). In this region, the comparison reveals a clearer separation between the results associated with the two chemical scenarios. Scenarios with negligible or moderate radial drift (e.g. 0.1-20 \si{\micro\meter}-sized grains) typically lead to higher ratios - C/O~$>1.5$ and N/O~$>4$ - under chemical inheritance, but yield lower values in the reset case, namely C/O~$<1.5$ and N/O ratios ranging from stellar to $4\times$stellar. The different trend is again a consequence of the greater fraction of oxygen that remains in the gas phase as O$_2$ after a chemical reset. Regions with a relatively high C/O* ratio ($>1.5$) and a relatively low N/O* ratio ($>4$) are instead only populated by evolutionary scenarios with efficient radial drift (e.g. 100\si{\micro\meter}-sized grains after 1Myr).

Between the CO$_2$ and CH$_4$ snowlines (Region 3, bottom left), the separation between the inheritance and reset scenarios becomes even clearer. In the reset scenarios, the evolutionary tracks remain limited to substellar C/O* ratios, with slightly superstellar values only reached in the scenario with 100~\si{\micro\meter}-sized grains after 1~Myr. In the inheritance scenarios, the C/O* ratio is always superstellar and ranges from about 1.3 to 2 for high ionisation levels and from 2 to 2.5 for low ionisation levels. The N/O* ratio spans a much larger dynamic range globally than in Region 2, ranging from substellar to strongly superstellar ($>10$) values, and is always lower in the reset scenarios than in the inheritance scenarios.

The steep increase in the N/O* ratio in Region 3, especially at later stages in the inheritance-high scenario, results from oxygen reaching a much lower abundance than nitrogen in the cold outer disc due to chemical processing and the relative positions of the snowlines of their main volatile carriers (see Figs.\ref{fig:molec_main} and \ref{fig:molec_nbear}). At larger distances, this effect becomes even more pronounced, leading to higher N/O* ratios. Beyond the CH$_4$ snowline (Region 4, bottom-right), the C/N* ratio provides a more robust diagnostic. In this region, comparing C/N* with C/O* helps resolve degeneracies in the chemical scenario and ionisation level that would otherwise arise from relying on a single elemental ratio. For example, although the C/N* ratio remains substellar across all scenarios, it exceeds 0.15 only in weakly ionised environments and remains below 0.10 under high ionisation. The C/O* ratio then allows for distinguishing between the inheritance scenario (C/O~$\gtrsim 1.7$) and the reset scenario (C/O~$\lesssim 1.7$).

\section{Discussion and comparison with previous works}

A key difference between our study and earlier models of coupled chemical and dynamical evolution in protoplanetary discs \citep[e.g.][]{Eistrup2018, Cridland2017b, Booth2019, Soto2022, Mah2023} lies in the treatment of the initial chemical conditions of volatiles (i.e. gas and ices). Earlier works typically assumed a uniform, subsolar volatile C/O ratio throughout the disc, consistent with the elemental composition of ices in dark clouds \citep[e.g.][]{Marboeuf2014, MummaCharnley2011}. Our compositional model additionally accounts for the fraction of oxygen sequestered in phyllosilicates and other O-bearing minerals, based on meteoritic constraints, as well as the presence of semi-refractory organic carbon, which contains a significant fraction of the total carbon budget and progressively sublimates in the inner disc (see Sect.~\ref{sec:comp_model}). The resulting elemental ratios for the volatile component (gas + ice mantles) as a function of radius and time are presented in App.~\ref{app:C}. The disc is initialised with a radial gradient in the volatile C/O ratio: subsolar ($\sim 0.35$) in the outer disc, where $C_{\rm ref}$ remains in the refractory phase, increasing interior to 5~au, and reaching supersolar values ($\sim 1.07$) inside 1~au, where all $C_{\rm ref}$ is in the gas. As a consequence, the disc chemistry evolves under a range of C/O regimes, including C-rich environments that promote the formation of hydrocarbons and oxygen-poor organics (see Sects.~\ref{sec:res_static_model} and \ref{sec:COratio}). Observational evidence for such C-rich chemistry has recently emerged from JWST observations of discs around low-mass stars \citep[e.g.][]{Tabone2023, Dishoeck2023, Colmenares2024}. Although this study focuses on a solar-mass star, our model provides a general theoretical framework for interpreting the origin of C-rich environments, particularly in scenarios involving radial drift and the erosion of refractory organic carbon in the inner disc. However, a direct comparison with observations will require extending the model to include the vertical disc structure and implementing full radiative transfer calculations.

A second key difference in our model is the inclusion of planetesimal formation, which is typically neglected in coupled chemical-dynamical models. By permanently removing volatiles from the cycling material, planetesimal formation alters the chemical regimes under which gas and dust evolve and affects the distribution of elemental ratios across the disc (see Sect.~\ref{sec:res_volatile_strong}). 

A consequence of the above considerations is that our simulations yield gas-phase C/O ratios that remain supersolar at all times within the $C_{\rm ref}$ snowline, across all analysed scenarios. In contrast, previous studies typically find subsolar C/O ratios in the inner disc, driven by the efficient release of oxygen from water-ice sublimation. For example, \citet{Soto2022}, who adopt the same chemical network as this work and model monodisperse dust populations, report C/O~$< 1$ within 10~au across both inheritance and reset scenarios. Comparable results are found by \citet{Booth2019} for the inheritance case. \citet{Mah2023}, who include a refractory carbon erosion front at $T > 631$~K, obtain supersolar C/O between the water and carbon snowlines only after 3~Myr in discs around solar-mass stars with $\alpha = 10^{-3}$.

This comparison highlights the sensitivity of coupled chemical-dynamical evolution to key model assumptions, including the adopted stellar abundances, the initial partitioning of elements into volatile, refractory, and semi-refractory phases, and the relative volatility of each component. In particular, the chemical nature and sublimation properties of semi-refractory species such as $C_{\rm ref}$, which play a key role in shaping the disc’s elemental budget, warrant further investigation to better constrain their influence on the chemical environments of planet-forming regions. 

\section{Caveats of model assumptions}\label{sec:caveats}

This study aims to investigate how key processes and initial conditions impact the chemical composition of protoplanetary discs in the regions where planet formation takes place. While our model accounts for the complex interplay between chemical evolution and mass transport, it is not intended to capture every aspect of disc physics and chemistry.

We treat the disc as an infinitely thin structure that evolves in isolation from its surrounding environment, neglecting effects such as late accretion of material onto the disc \citep[e.g.][]{Gupta2023}, external feedback from nearby stars \citep[e.g.][]{Keyte2025}, and mass loss via disc winds \citep[e.g.][]{Lesur2021}. These processes can influence the disc's density and temperature structure, with potential chemical implications, and should be included in more comprehensive models of disc evolution.

In addition, our model considers the midplane to be chemically isolated from the upper disc layers. Vertical mixing and large-scale meridional flows can redistribute material processed in the warm, UV- and X-ray-irradiated surface regions down to the midplane, potentially introducing additional chemical pathways not captured here \citep[e.g.][]{Semenov2011, Furuya2014}. Including such effects would require coupling the vertical disc structure with radiative transfer and dynamical mixing, which lies beyond the scope of this work.

At the start of our simulations, we assume that the dust has fully settled in the midplane, with an initial dust-to-gas mass ratio of 0.01. We further assume that the formation of planetesimals occurs early ($t\sim10^5$~yr) and with uniform efficiency across the dust disc, converting $50\%$ of the initial dust into large bodies that are chemically decoupled from the gas and the remaining submillimetre grains. 

Evidence from the Solar System supports the assumption that the radial extent of the planetesimal disc matches that of the dust disc \citep{Kretke2012}. However, the primary mechanisms governing planetesimal formation and their radial efficiency remain uncertain (see \citealp{Drazkowska2023} for a recent review). The depletion of volatiles in the inner disc due to mass transport following planetesimal formation (Sect.~\ref{sec:res_volatile_strong}) could be more pronounced if a larger fraction of the initial dust were converted into planetesimals or less pronounced if planetesimal formation were less efficient in the outer disc (e.g. beyond the CO$_2$ snowline). 

The timing of planetesimal formation also remains an open question. Planetesimals must form early enough for giant planets to grow before the gas in the disc disperses, typically within $\sim5-10$~Myr. The process can begin as early as the Class 0/I stage. For example, \citet{Bonsor2023} characterised planetary material recently accreted by white dwarfs, whose spectra suggest the accretion of iron-core or mantle-rich material, which can only form in the presence of strong heating, most plausibly from the decay of SLRs with half-lives less than 1~Myr. This points to planetesimal formation occurring as early as $\sim 10^5$ years after cloud collapse (see also \citealp{Cridland2022}; \citealp{Huhn2025}). Protoplanetary discs may therefore enter the Class II stage with a significant fraction of their dust already converted into planetesimals. This hypothesis is further supported by models suggesting a significant dust drift prior to disc formation, potentially leading to a dust-enriched disc at early times \citep{Lebreuilly2020}. Observationally, dust masses in Class II discs are often significantly lower than those inferred for embedded Class 0/I systems \citep[e.g.][]{Tychoniec2020}, consistent with the idea that a considerable amount of dust may already have been incorporated into larger bodies by the time discs reach the Class II stage. If the process of planetesimal formation introduces radial inhomogeneities in the volatile content of the remaining dust, then our results suggest that it should be modelled alongside chemical processes.

For the residual dust grains, each simulation assumes a single size in the sub-mm regime, neglecting the effects of dust coagulation and fragmentation. We recognise that, in a more realistic model, these processes lead to a continuous size distribution in which large grains (near the fragmentation barrier) dominate the mass, while small grains (resulting from fragmentation) dominate the surface area available for chemistry \citep[e.g.][]{Birnstiel2024}. Allowing grains to grow beyond 100~\si{\micro\meter} would reduce the efficiency of surface chemistry, leading to slower variations in ice-phase abundances due to chemical reactions \citep{Eistrup2022, Navarro2024}. At the same time, larger grains drift faster, so the general trend of sublimation-driven enrichment of the gas phase would likely be preserved or even enhanced. In addition, if grain growth significantly reduces the mass fraction of the smallest grains, our results may overestimate their chemical contribution. Alternatively, our small-grain results may be representative of discs where abundant small grains persist due to the presence of second-generation dust produced by collisions of larger bodies \citep{Turrini2019, Guidi2022}. Future work incorporating a full grain-size distribution and exploring different scenarios for planetesimal formation will be required to resolve these degeneracies and quantify the effects more precisely.

\section{Summary and conclusions}\label{sec:concl}

In this study, we investigated how chemical processes and mass transport in the midplane of a Class II protoplanetary disc shape the composition of the gas and dust from which planets form. Specifically, we examined the combined effects of volatile chemical processing, the viscous evolution of the gas, and the radial drift of the dust, taking into account both gas-phase and surface chemistry as well as viscous heating and turbulent mixing.

While most studies focus on discs that retain their initial molecular inventory, we also investigated the impact of alternative scenarios for the volatile chemistry. Specifically, in addition to the full-inheritance scenario, we also studied the case of a complete reset of the chemistry and analysed both scenarios under two ionisation environments: one dominated by the decay of short-lived radionuclides and another with an additional contribution from cosmic rays. We derived the initial budget of volatiles in the disc in the context of a more realistic and observationally constrained compositional model that accounts for the presence of refractory and semi-refractory components, including refractory organic carbon. In addition, we analysed three different scenarios for the size of the dust grains. We focused on submillimetre grains (0.1, 20, and 100~\si{\micro\meter}), i.e., grain sizes significantly smaller than those typically assumed for pebbles in discs (mm-cm), and quantified the impact of processes such as ice sublimation due to such a population of relatively small grains.

We simulated the evolution of the disc over 3~Myr using our fully integrated 1D code {\sc{JADE}}, assuming a viscous $\alpha$ parameter of $10^{-3}$. The analysis of the resulting time-dependent molecular abundance profiles and elemental ratios (C/O, C/N, and N/O) in the disc reveals that: 

\begin{itemize}

    \item Chemistry and dynamics in discs are tightly interconnected and must be modelled simultaneously. This applies even in discs where dust remains strongly coupled to the gas (grain size $\lesssim 0.1$~\si{\micro\meter}) and radial drift is negligible. Radial diffusion and mixing, in particular, tend to smooth out local peaks in the abundances of volatile species, which often emerge in chemically active environments (e.g. after a chemical reset or under high ionisation conditions). 

    \item Radial drift can drive significant volatile enrichment in the gas phase, even for relatively small grains ($\lesssim 100$~\si{\micro\meter}), on timescales comparable to those of planet formation. Within their snowlines, species such as H$_2$O, CO$_2$, and NH$_3$ can reach enrichment factors of nearly twofold for 20~\si{\micro\meter}-sized grains and almost an order of magnitude for 100~\si{\micro\meter}-sized grains compared to the 0.1~\si{\micro\meter} case over 3~Myr. In contrast, highly volatile species such as CO and N$_2$ experience only modest enrichment. These results suggest that giant planets can accrete metal-rich gas from sublimated icy grains, even when the bulk of pebbles (mm–cm-sized dust grains) have already been converted into larger bodies or accreted onto the star, leaving only smaller grains in the disc.

    \item The formation of planetesimals in a viscously evolving disc can indirectly alter the volatile budget and metallicity of the gas phase within the snowlines. This effect results from the combined processes of inward mass transport and ice sublimation. Planetesimals sequester part of the initial ice budget in a chemically inert reservoir, effectively reducing the total amount of volatiles carried by dust grains. As the disc evolves, the gas within the snowlines is replaced by gas from the outer disc, which is enriched by the sublimation of the volatile-depleted grains. If planetesimals form early and with constant efficiency throughout the disc, this process leads to a net depletion of volatiles in the inner-disc gas phase on short timescales ($\lesssim 0.5$~Myr). This depletion exhibits a strong volatility-dependent trend: species with a lower volatility, such as H$_2$O or CO$_2$, are more depleted than highly volatile species such as N$_2$, which in turn also changes the gas-phase elemental ratios. These results underline the importance of accounting for planetesimal formation in models that aim to link disc chemistry to planetary compositions. 
    
    \item Disc evolution leads to strong local enhancements in the C/O* ratio, up to $8\times$ stellar in the gas phase (C/O~$\sim 4.72$ on the solar scale) and $2\times$ stellar in the solid phase (C/O~$\sim 1.2$ on the solar scale) in models with significant radial drift (e.g. 100~\si{\micro\meter}-sized grains). Even in the less perturbed scenario - inheritance with ionisation from SLRs and 0.1~\si{\micro\meter}-sized grains - temporal variations range within $20\%$ of the stellar value, with peaks of $50\%$ between 1.5 and 3~au. Notably, none of the analysed scenarios reproduce the classical monotonic trend of the gas-phase C/O ratio with radial distance \citep{Oberg2011}, which is widely assumed in planet formation studies.

    \item The erosion of refractory organic carbon can significantly influence the composition of the gas in the inner disc. If released as atomic carbon between the H$_2$O and CO$_2$ snowlines, it reacts to form carbon chains and organic compounds that subsequently condense over a timescale of about 0.5~Myr. This process typically leads to a net decrease in gas-phase carbon abundance. However, in the case of strong radial drift, carbon is enriched in the gas phase more efficiently than it is reprocessed, resulting in a net increase. This mechanism also affects the elemental ratios, driving local enhancements of up to $4.5\times$ stellar in the C/O* ratio and $3.7\times$ stellar in the C/N* ratio. These results highlight the potentially critical role of refractory organic carbon on the chemical composition of planetary building blocks, emphasising the need to account for this additional carbon reservoir in planet formation models. In this respect, efforts aimed at characterising the mechanisms behind carbon erosion in the inner region of protoplanetary discs are highly valued, and further investigations are strongly encouraged. 

    \item While individual elemental ratios provide valuable insights into disc evolution, comparing multiple ratios is key to isolating signatures of specific scenarios. For example, a pairwise comparison of the C/O* and N/O* ratios between the H$_2$O and CH$_4$ snowlines reveals distinctions between the inheritance and reset scenarios that would be less clear when considering either ratio alone. Beyond the CH$_4$ snowline, comparing the C/O* and C/N* ratios further constrains the ionisation level.

\end{itemize} 

\noindent Our results demonstrate that chemical processes, combined with gas and dust transport, significantly alter the composition of protoplanetary discs over timescales exceeding 2~Myr and comparable to those of planet formation. Disc evolution cannot be neglected, even in systems that have retained their original molecular inventory or where dust grains experience minimal radial drift. Consequently, the growth and migration of planetary bodies must be modelled alongside the chemical and dynamical evolution of their surrounding environment. In a forthcoming companion paper, we will investigate how this evolving disc composition influences the chemical makeup of giant planet atmospheres.

\begin{acknowledgements}
    This work was carried out within the framework of, and with the support of, the European Research Council via the Horizon 2020 Framework Programme ERC Synergy “ECOGAL” Project GA-855130. E.P. and D.T. acknowledge support from the Italian Space Agency (ASI) through the ASI-INAF grant No.2016-23-H.0 plus addendum No. 2016-23-H.2-2021, and the ASI-INAF contract No.2021-5-HH.0. C.W.~acknowledges financial support from the Science and Technology Facilities Council and UK Research and Innovation (grant numbers ST/X001016/1 and MR/T040726/1). The computational resources for this work were supplied by the Genesis cluster at INAF-IAPS and the technical support of Scigé John Liu is gratefully acknowledged. Finally, we thank the anonymous referee for their constructive comments and suggestions, which helped improve the clarity and quality of this manuscript.
\end{acknowledgements}

\bibliography{references.bib}
\bibliographystyle{aa}

\begin{appendix} 
\section{Impact of planetesimal formation}\label{app:A}

\begin{figure*}
\centering
\includegraphics[width= \textwidth]{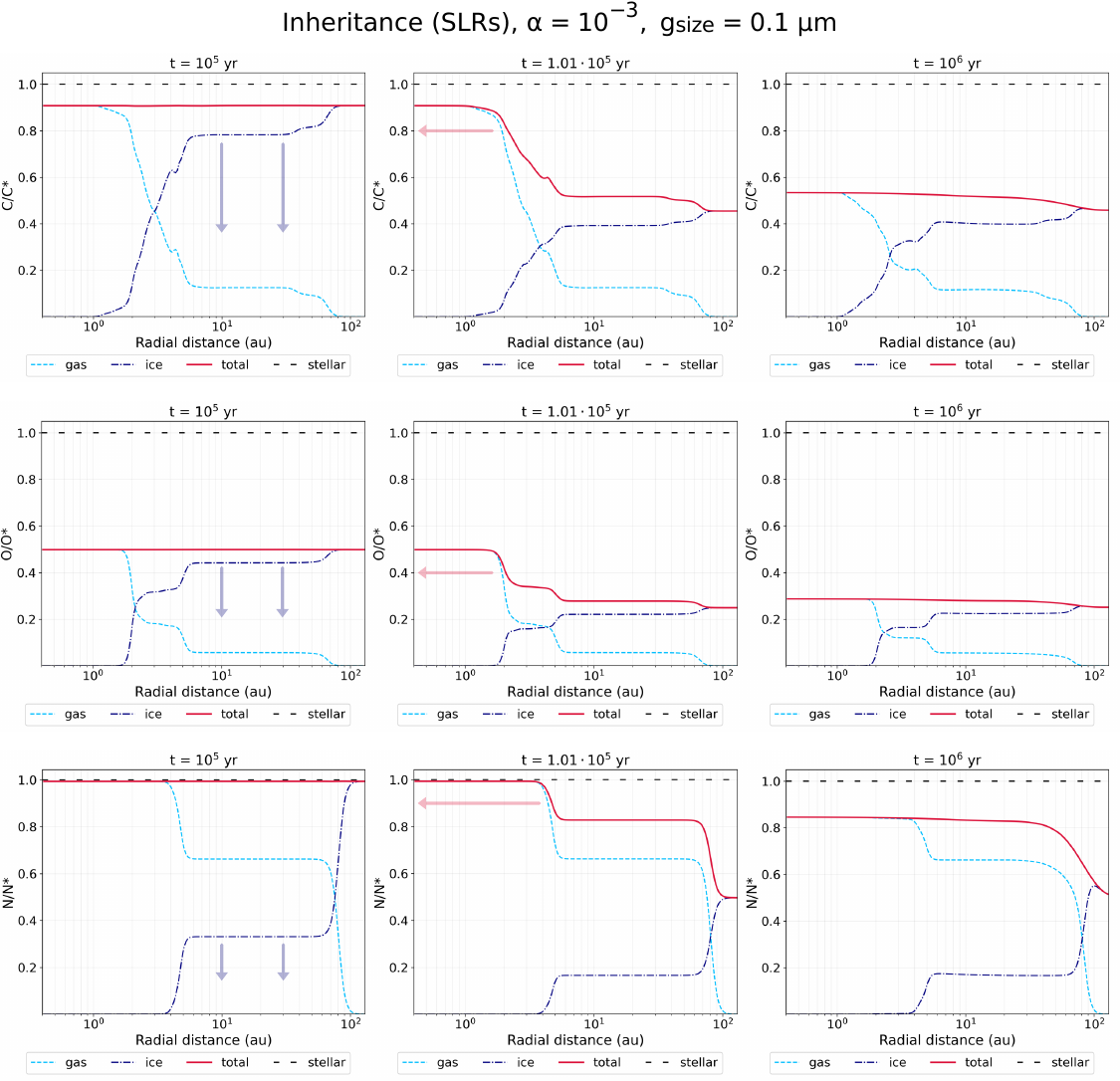}
\caption{Radial abundance profiles of C, O, and N in the disc at the point of planetesimal formation ($10^{5}$~yr), $10^3$ years later, and at 1~Myr in the inheritance-low scenario with 0.1~\si{\micro\meter}-sized grains. The light blue curve (dashed) refers to the gas phase, while the blue curve (dash-dotted) indicates the ice phase. The red curve (solid) is the sum of the two contributions, initially equal to the protosolar abundance of each element, reduced by the meteoritic abundance. The plot shows the evolution of the abundances in the disc due to mass transport during viscous evolution.} 
\label{fig:plts}
\end{figure*}

In this appendix, we examine how the composition of the disc changes due to mass transport after the conversion of half of the dust into planetesimals. We assume that this conversion occurs with the same efficiency throughout the disc and is already completed $10^5$~yr after the zero time we set for a Class II disc in our simulations. For simplicity, we assume that the composition of the planetesimals is fixed by the composition of the solid phase (refractories and ices) at $10^5$~yr. The evolution of the total elemental abundances in the disc after $10^5$~yr is analysed in detail in Fig.~\ref{fig:plts} for the inheritance-low scenario with 0.1~\si{\micro\meter}-sized grains. Each panel shows the radial abundance profiles of the volatile fractions of C, O, and N in the gas phase (dashed lines) and as ice on the surface of the dust grains (dash-dotted lines) at three times: $10^{5}$~yr, $10^{3}$~years later, and after 1~Myr of evolution of the disc with {\sc JADE}.

At $10^{5}$~yr, the sum of the gas and ice abundances (solid red curve) is constant in the entire disc and results from the difference between the protosolar and meteoritic abundance for each element. After the formation of planetesimals, which effectively represent a chemically inert reservoir of volatile and refractory elements, the abundance of volatile elements that remain in the disc as ice decreases by a factor of 2. Beyond the snowlines, the total abundance of elements in the gas and on the dust is therefore lower than in the inner disc, where all volatile elements are in the gas phase and planetesimals only incorporate refractory elements (centre column in Fig.~\ref{fig:plts}). If there were no viscous evolution, the discontinuities in total abundance would remain localised at the snowlines. However, viscous evolution leads to mass accretion onto the star and transport of dust grains, together with their ice mantles, from the outer to the inner disc. 

\begin{figure*}
\centering
\includegraphics[width= \textwidth]{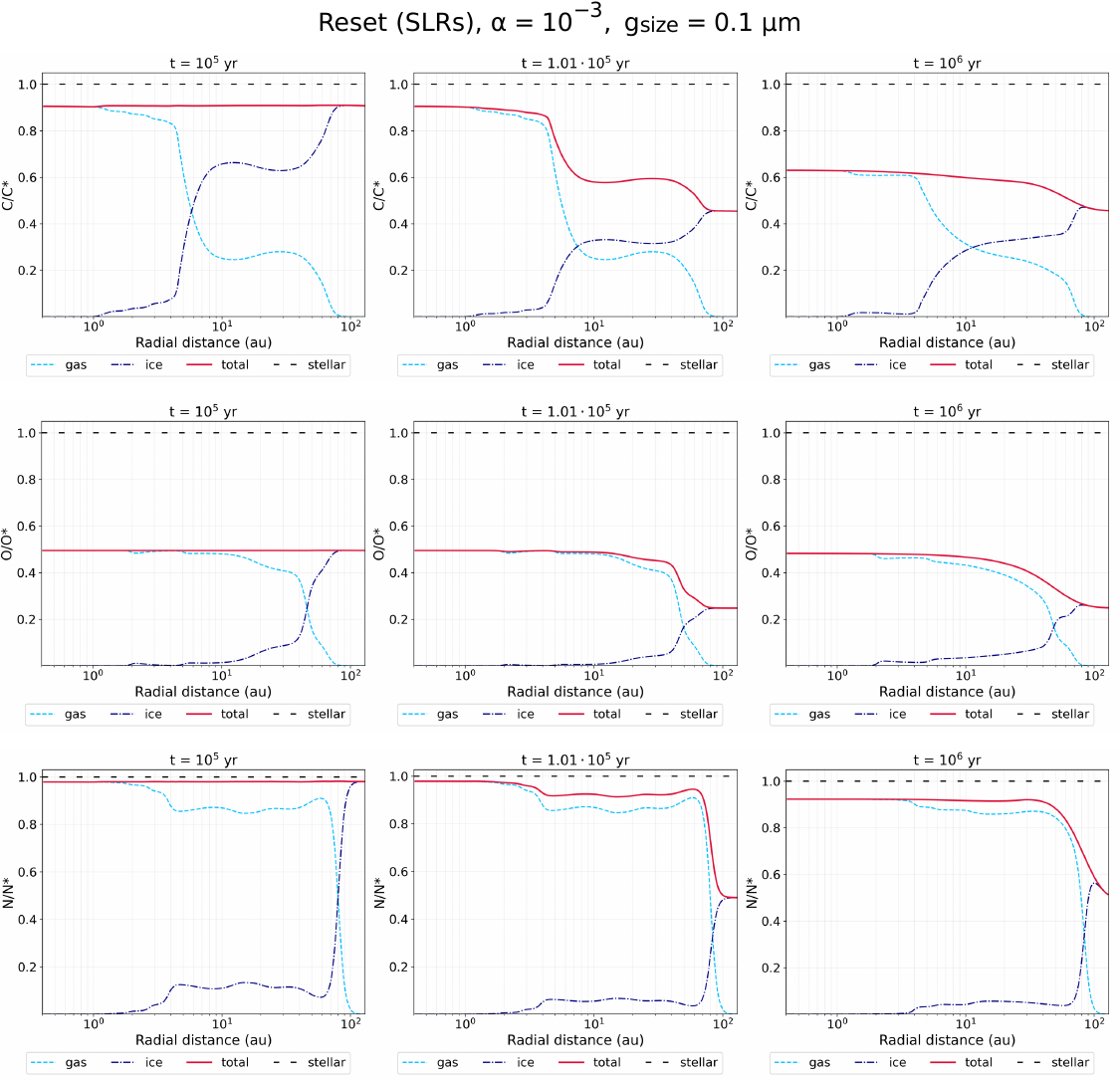}
\caption{Same as Fig.~\ref{fig:plts} for the reset-low scenario.} 
\label{fig:plts_rl}
\end{figure*}

Focusing on nitrogen, the total mass of the disc gas within the snowline of NH$_3$ ($r\sim4.5$~au) is about $6.5\times10^{-4} M_\odot$, which corresponds to about $1.2\%$ of the total mass of the disc in our model. This mass is accreted to the star in the first $4\times10^{5}$~yr (given a mass accretion rate of $\sim10^{-9} M_\odot/{\rm yr}$). On this timescale, the region within the NH$_3$ snowline is thus emptied and replenished with material from the outer disc, which has a lower N abundance in the ice phase. The icy grains that cross the snowline release their N back into the gas phase, which is then depleted in N compared to the initial state. On a timescale of 1~Myr, the total abundance of volatile N in the inner disc gradually settles to the value reached in the outer disc after the formation of planetesimals (lower-right panel in Fig.~\ref{fig:plts}). The decrease in N gas abundance in the inner disc continues until the icy material from beyond the N$_2$ snowline is also transported inwards, depleting the inner gas phase by even more N$_2$. However, this happens on long timescales that are not captured by our simulations. The effect described above explains why the gas in the inner disc is more depleted in NH$_3$ than in N$_2$ (see the molecular abundance profiles in Fig.~\ref{fig:molec_nbear}). Similar arguments also apply to the molecular carriers of C and O and explain the general decrease in their abundance in the gas and ice phase after 3~Myr compared to the initial values (Fig.~\ref{fig:molec_main}). A direct consequence of this process is that the viscosity-driven physical evolution of the disc can lead to substellar compositions of the gas in the inner disc, independent of chemical processes. The same analysis is shown in Fig.~\ref{fig:plts_rl} for the reset-low scenario. 

\section{Elemental ratios at high ionisation}\label{app:B}

In this appendix, we present the radial and temporal evolution of the gas- and solid-phase C/O, C/N, and N/O ratios in the inheritance and reset scenarios under high ionisation, for all three grain size cases: 0.1, 20, and 100~\si{\micro\meter}. Elemental abundances are derived from the molecular abundance profiles returned by each simulation, considering only species with abundances above $10^{-7}$ relative to total hydrogen nuclei. Solid-phase ratios include contributions from rocks, ice mantles, and refractory organic carbon. All elemental ratios are normalised to their respective stellar values.

\begin{figure*}
\centering
\begin{subfigure}[b]{\textwidth}
    \centering
    \includegraphics[width=0.95\textwidth]{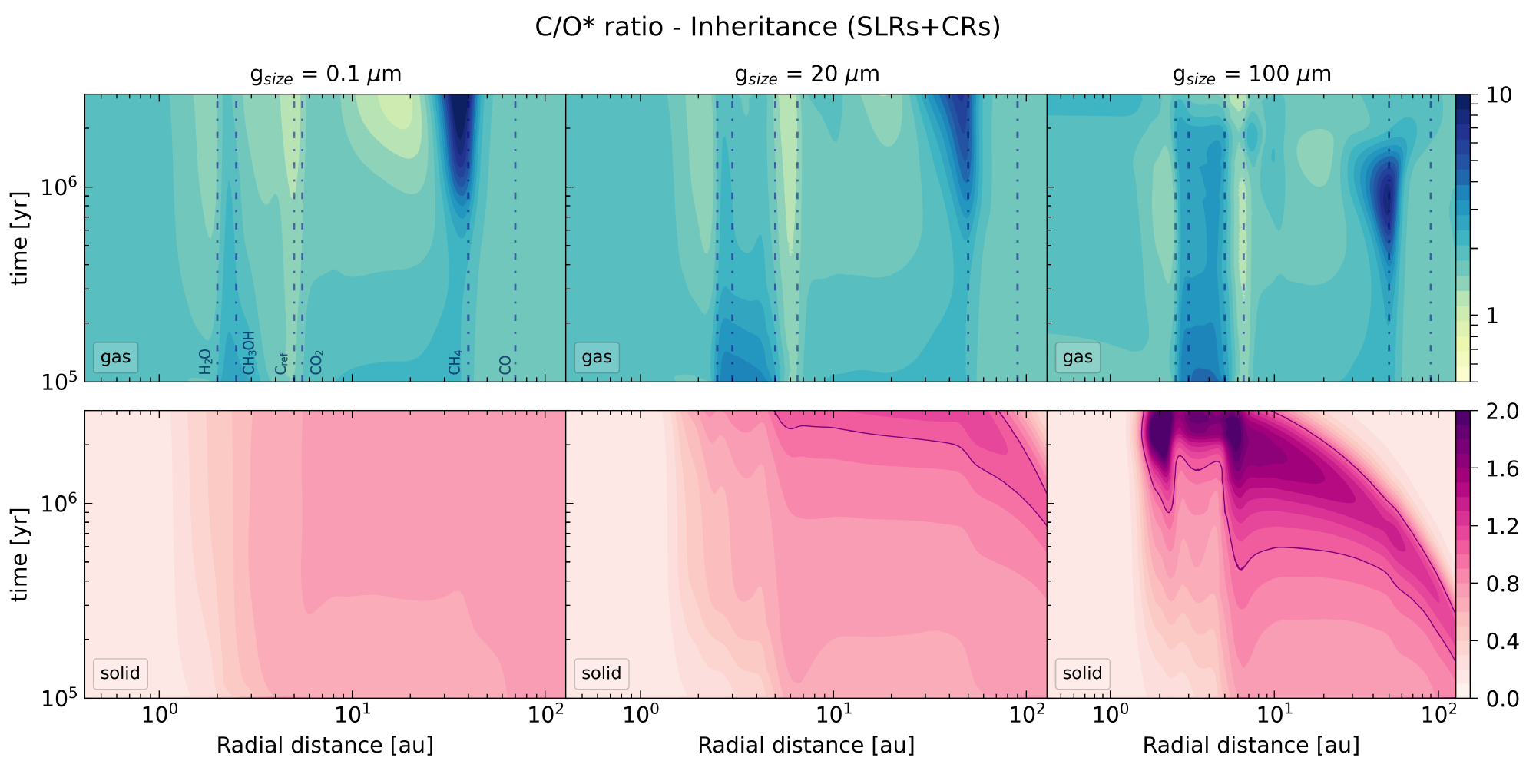}
    \caption{}
\end{subfigure}
\vspace{2mm} 
\begin{subfigure}[b]{\textwidth}
    \centering
    \includegraphics[width=0.95\textwidth]{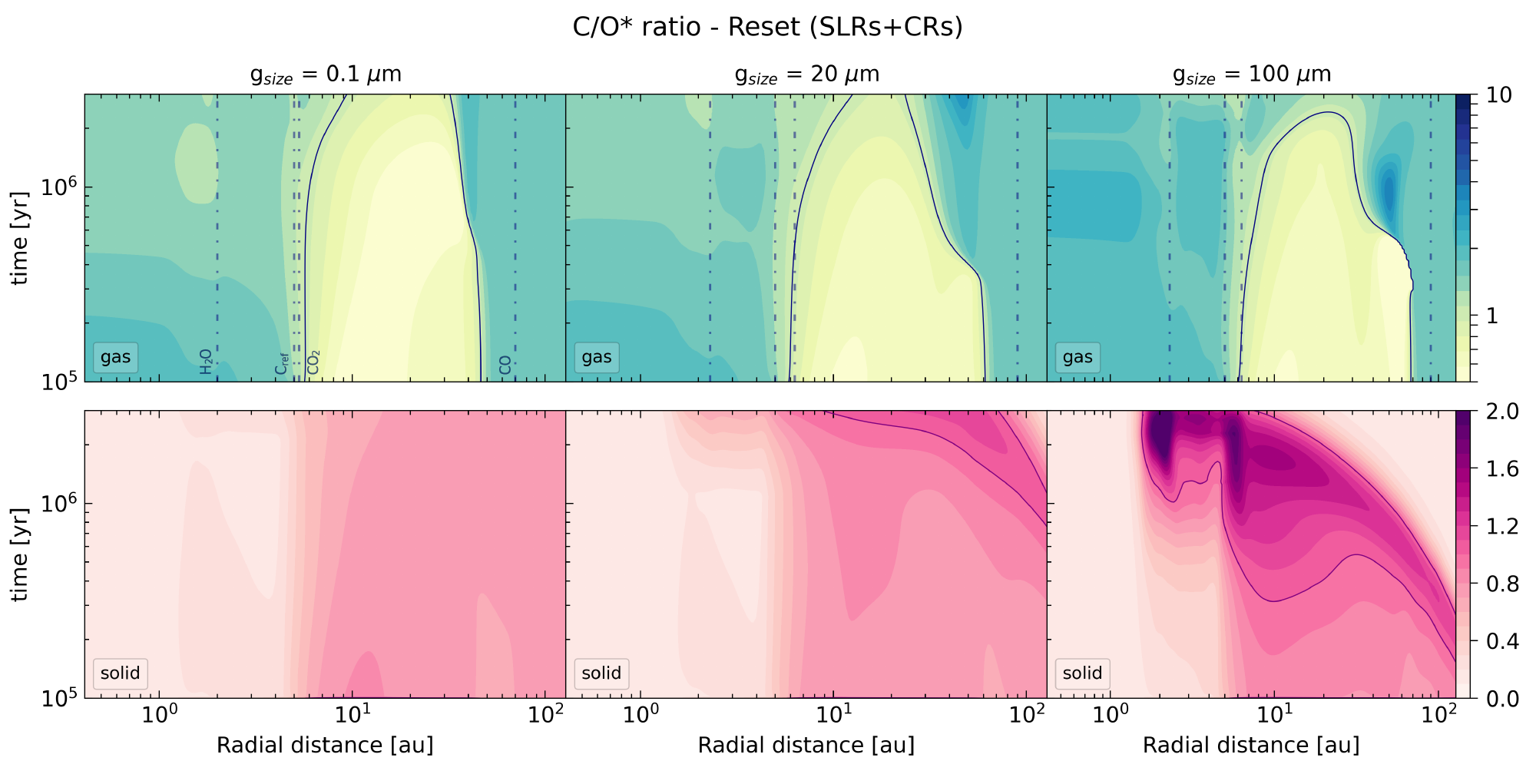}
    \caption{}
\end{subfigure}
\caption{Same as Fig.~\ref{fig:c_o_ratio} for the inheritance (panel \emph{a}) and reset (panel \emph{b}) scenarios with high ionisation.}
\label{fig:c_o_ratio_high}
\end{figure*}

\begin{figure*}
\centering
\begin{subfigure}[b]{\textwidth}
    \centering
    \includegraphics[width=0.95\textwidth]{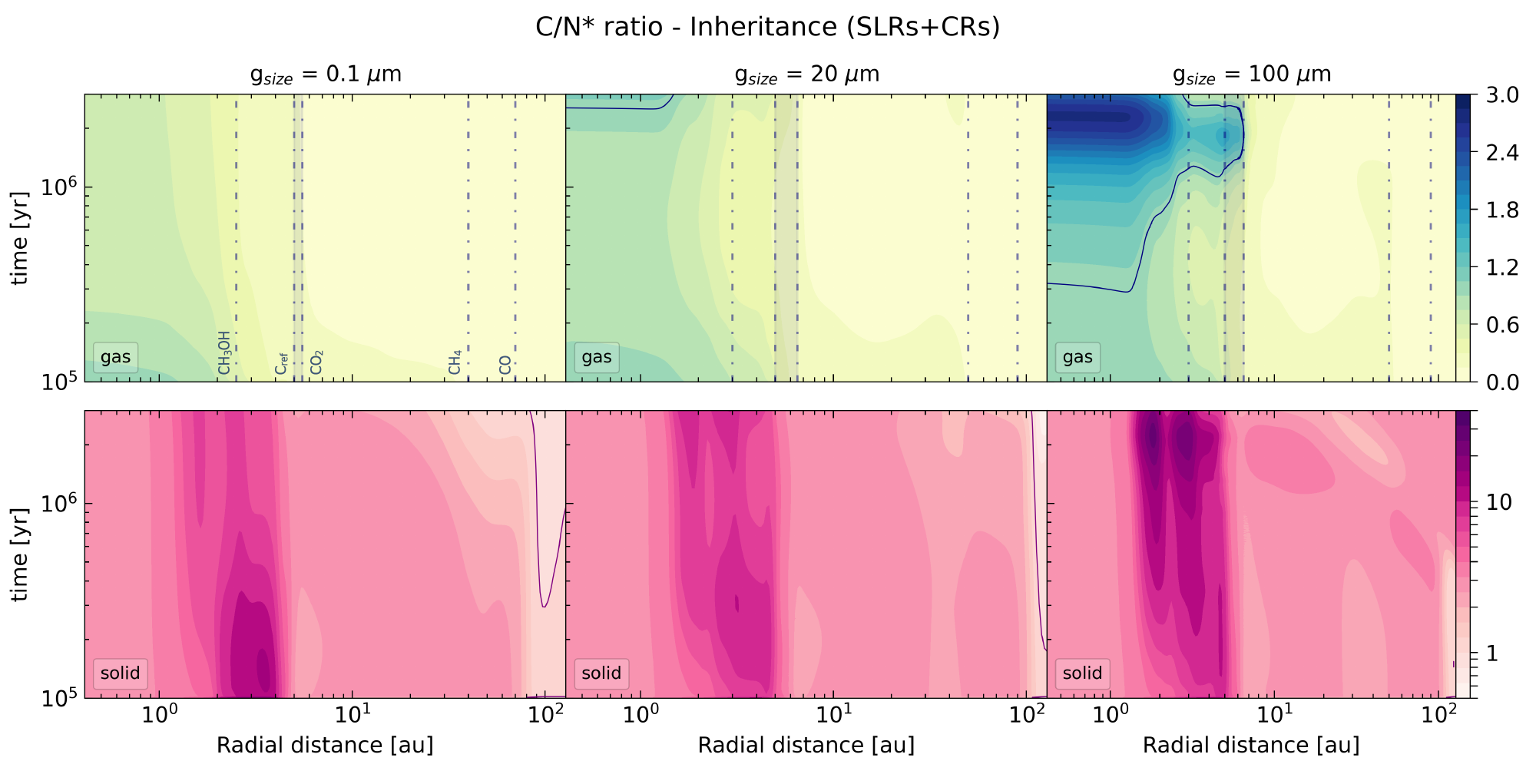}
    \caption{}
\end{subfigure}
\vspace{2mm} 
\begin{subfigure}[b]{\textwidth}
    \centering
    \includegraphics[width=0.95\textwidth]{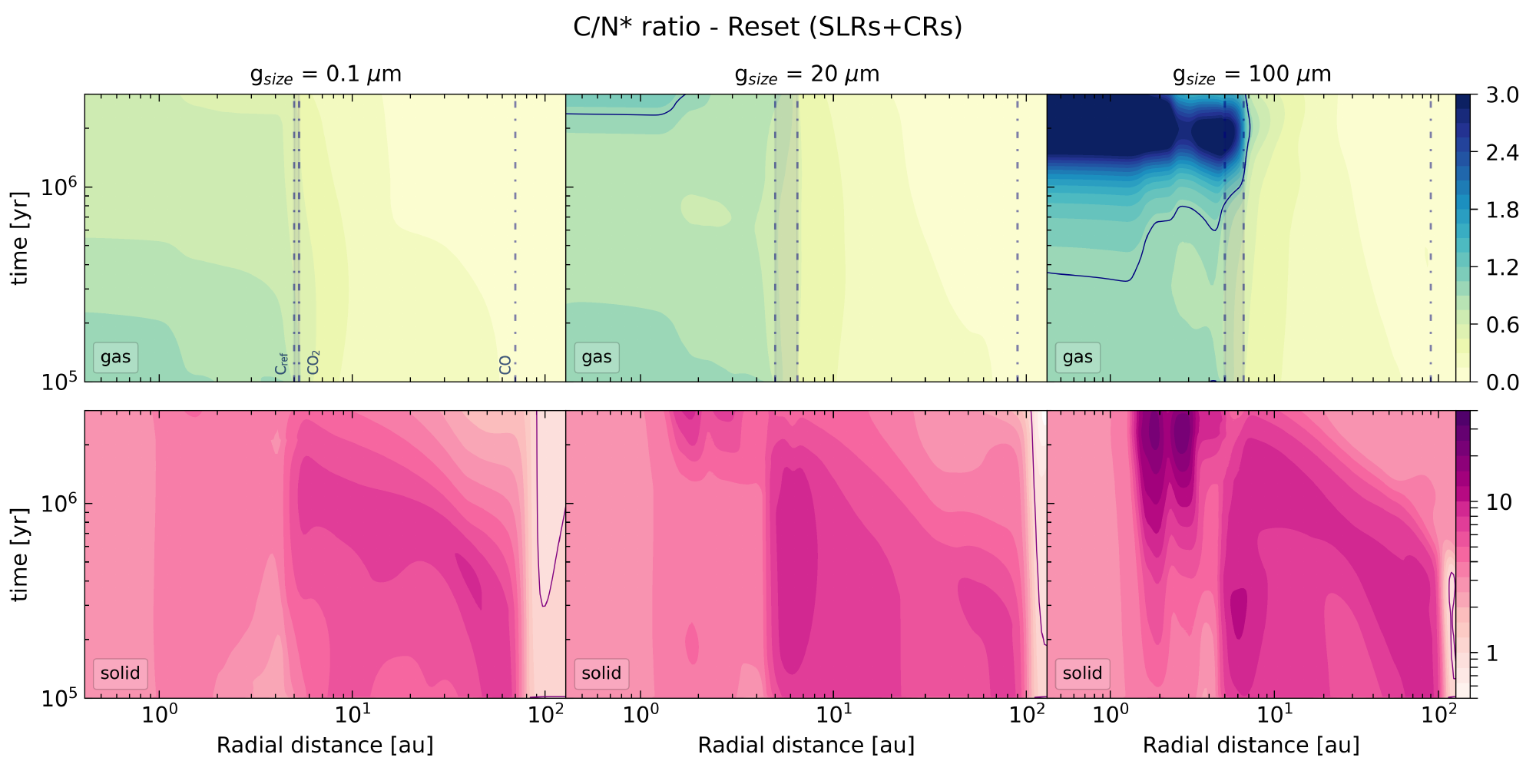}
    \caption{}
\end{subfigure}
\caption{Same as Fig.~\ref{fig:c_n_ratio} for the inheritance (panel \emph{a}) and reset (panel \emph{b}) scenarios with high ionisation.}
\label{fig:c_n_ratio_high}
\end{figure*}

\begin{figure*}
\centering
\begin{subfigure}[b]{\textwidth}
    \centering
    \includegraphics[width=0.95\textwidth]{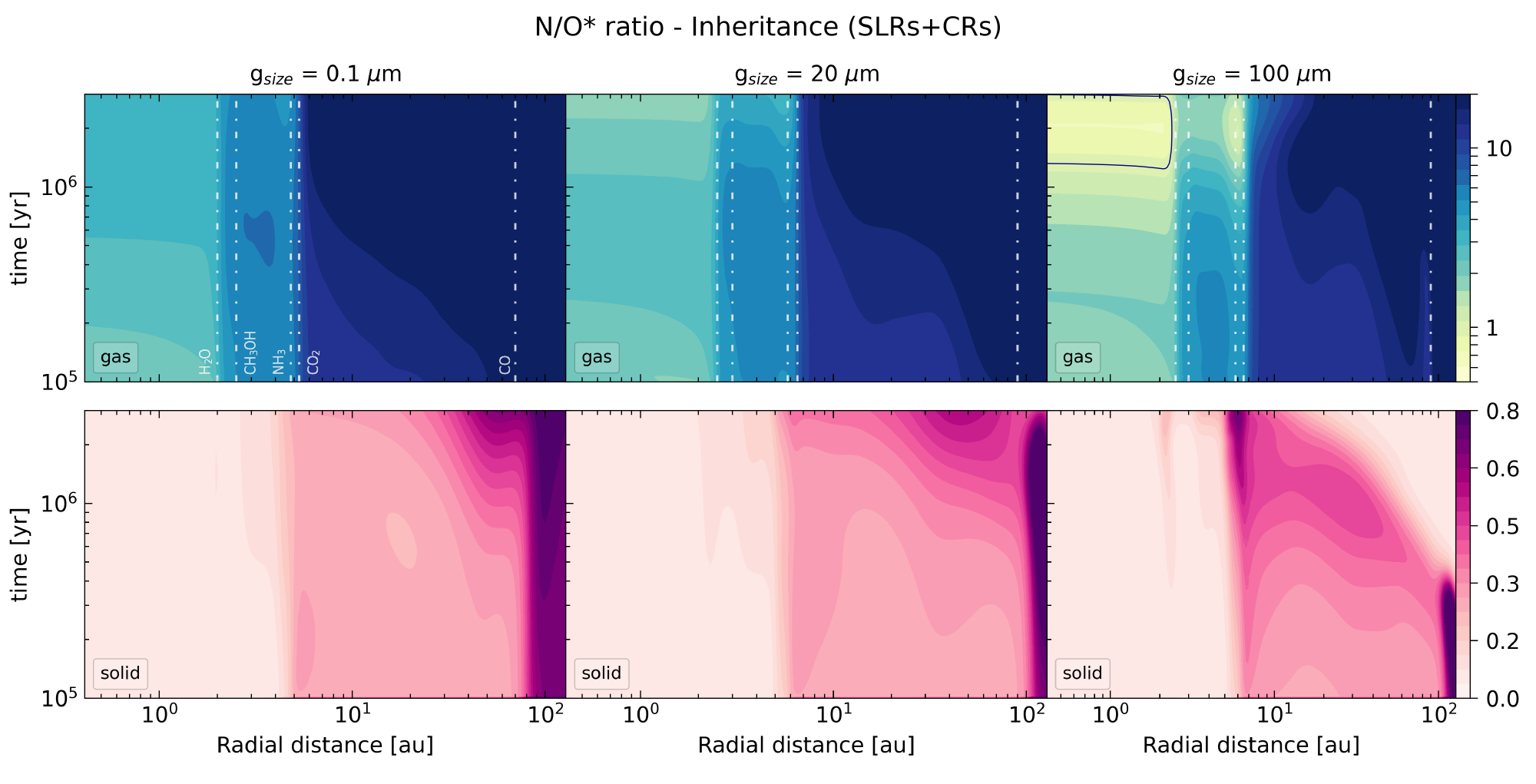}
    \caption{}
\end{subfigure}
\vspace{2mm} 
\begin{subfigure}[b]{\textwidth}
    \centering
    \includegraphics[width=0.95\textwidth]{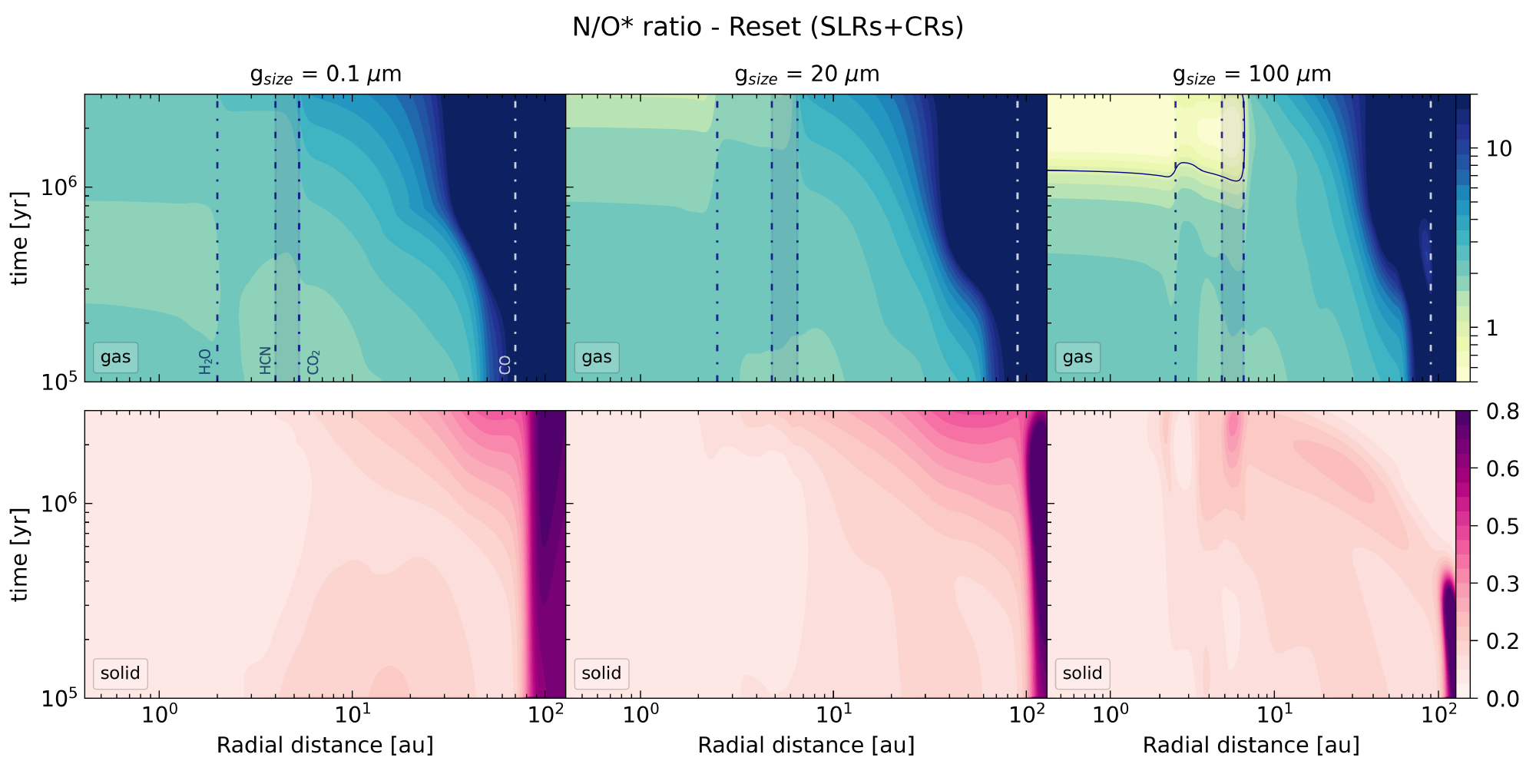}
    \caption{}
\end{subfigure}
\caption{Same as Fig.~\ref{fig:n_o_ratio} for the inheritance (panel \emph{a}) and reset (panel \emph{b}) scenarios with high ionisation.}
\label{fig:n_o_ratio_high}
\end{figure*}

\section{Total elemental ratios of volatiles}\label{app:C}

In this appendix, we present the radial and temporal evolution of the total elemental ratios of the volatile component, defined here as the sum of gas and ice mantles, for the three grain size scenarios in the inheritance case with low ionisation. The evolution is shown from $10^5$~yr onwards, that is, after the onset of planetesimal formation, which removes solids from the disc. The plotted ratios are normalised to their respective stellar values, and the dashed black contour indicates where the ratio equals the stellar value. As a consequence of the initial distribution of elements across volatiles, rocks, and semi-refractory organic carbon, the disc starts with radial gradients in the volatile elemental ratios. The volatile C/O ratio is initially substellar in the outer disc and becomes superstellar inside 5~au, where semi-refractory organic carbon ($C_{\rm ref}$) transitions to the gas phase. The C/N ratio closely follows the carbon distribution due to the high volatility of nitrogen: it is strongly substellar in the outer disc and increases towards the inner disc, becoming superstellar only in scenarios with efficient radial drift (100~\si{\micro\meter}-sized grains) after $\sim 0.4$~Myr. The N/O ratio is initially superstellar across the disc due to the sequestration of a large fraction of oxygen in phyllosilicates and other refractory minerals, and becomes substellar only in the 100~\si{\micro\meter} grain scenario after $\sim 1$~Myr, within the H$_2$O snowline.

\begin{figure*}
\centering
\includegraphics[width=0.95\textwidth]{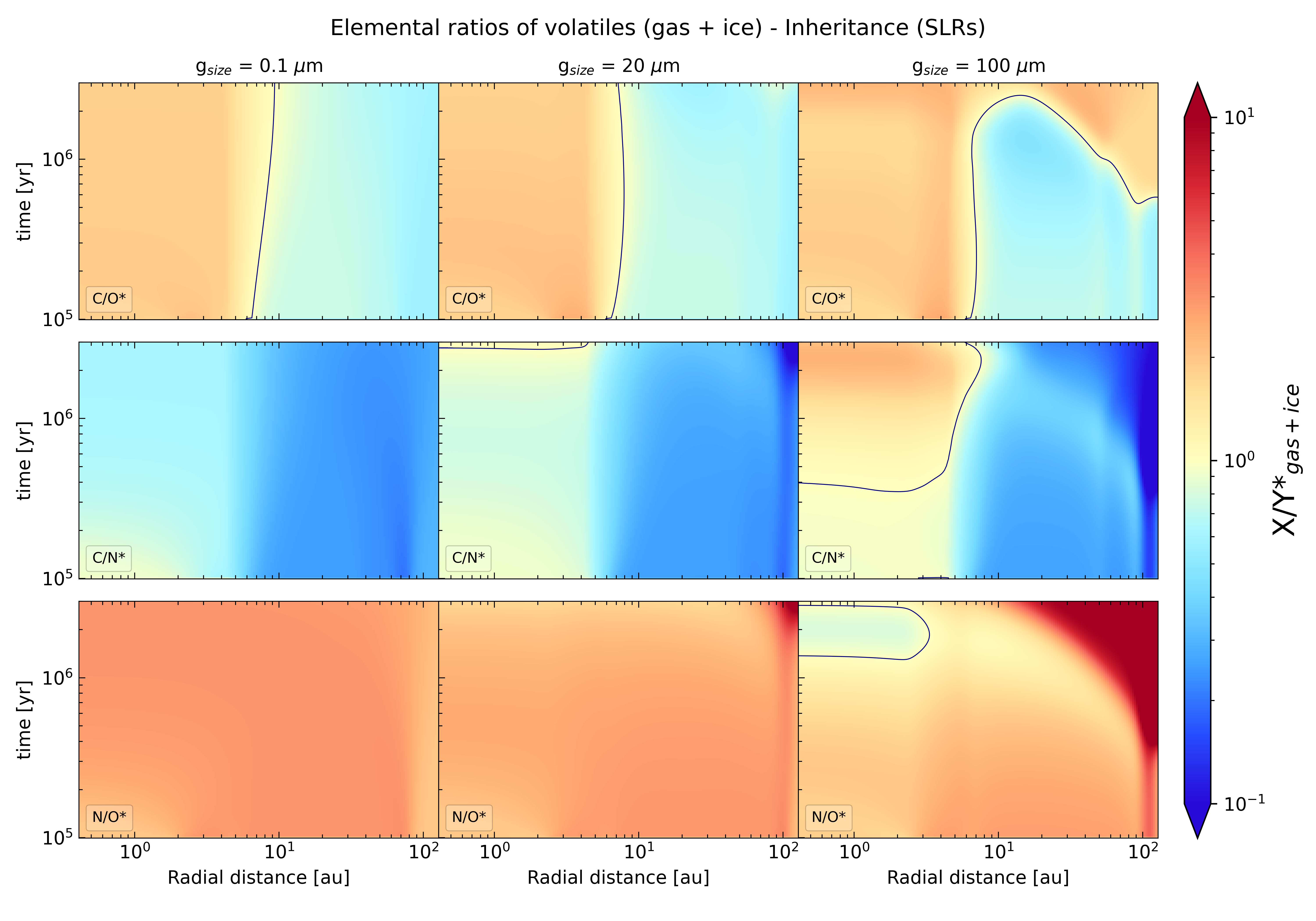}
\caption{Total elemental ratios of the volatile component (i.e. including only gas and ice mantles) as a function of radial position and time in the disc, for the three considered grain sizes in the inheritance scenario with low ionisation. All values are normalised to the respective stellar ratio, with the normalisation indicated by the superscript $^*$. The dark contour represents a value of~1, corresponding to the stellar ratio.} 
\label{fig:total_ratios}
\end{figure*}

\end{appendix}

\end{document}